\documentclass[acmsmall,screen,nonacm]{acmart}
\acmJournal{TACO}
\setcopyright{none}
\hypersetup{keeppdfinfo}
\usepackage{amsmath}
\usepackage{booktabs}
\usepackage{tabularx}
\usepackage{longtable}
\usepackage{array}
\usepackage{graphicx}
\usepackage{microtype}
\usepackage{placeins}
\usepackage{flafter}
\AtBeginEnvironment{thebibliography}{\interlinepenalty=10000\relax}
\usepackage{float}
\floatstyle{ruled}
\newfloat{algorithm}{tbp}{loa}
\floatname{algorithm}{Algorithm}
\newcolumntype{P}[1]{>{\raggedright\arraybackslash}p{#1}}
\newcolumntype{Y}{>{\raggedright\arraybackslash}X}
\newcommand{\Simple}{\textsc{Simple}}
\newcommand{\Local}{\textsc{Local}}

\newcommand{\Cfour}{\textsc{C4}}
\newcommand{\Pfour}{\textsc{P4}}
\newcommand{\TO}{\textsc{TO}}
\newcommand{\SD}{\textsc{SD}}
\newcommand{\biteq}{\mathrel{\overset{\mathrm{bits}}{=}}}
\newcommand{\paperfigure}[5][tbp]{\begin{figure}[#1]\centering\includegraphics[width=\linewidth]{figures/#2.pdf}\caption{#3}\Description{#4}\label{#5}\end{figure}}

\begin{document}
% Keep page-bottom slack at the bottom, rather than stretching float-to-text gaps.
\raggedbottom
\title[Recovering Parallelism under Fixed Numerical References]{Recovering Parallelism under Fixed Numerical References in Quantized LLM Decoders}
\author{Chengzhan Li}
\email{chengzhan.li@std.uestc.edu.cn}
\affiliation{\institution{University of Electronic Science and Technology of China}\city{Chengdu}\country{China}}
\authorsaddresses{Author's Contact Information: Chengzhan Li,
\href{mailto:chengzhan.li@std.uestc.edu.cn}{chengzhan.li@std.uestc.edu.cn}.\\
University of Electronic Science and Technology of China, Chengdu, China}
\renewcommand{\shortauthors}{C. Li}
\begin{abstract}
Splitting a quantized decoder at a physical attention-view boundary can break projection groups that already match a fixed numerical reference. LOCAL retains those groups, partitions queries by physical view inside attention, and rejoins the outputs in query order. Owned score rows survive native-buffer reuse, and commit supplies the key--value (KV) and consumer state for the next call. At a fixed 32B attention-view crossing on A5000 with frozen proposals, LOCAL reduces verification-through-commit time by 21.56\% against SIMPLE (the whole-call splitting baseline). This supporting endpoint covers 12 paired time blocks (95\% CI [20.21\%, 22.83\%]). In the primary noncrossing history, both arms retain the four-column projection and the estimated reduction is 0.43\% [$-1.13\%$, $2.01\%$]. This comparison measures the two implementations when neither fragments the projection group. Across two fixed 8B conversations under target-only decoding, the primary equal-weight estimate was a 1.44\% reduction in response-plus-reset time (95\% CI [$-0.82\%$, $3.85\%$]). Separate studies with SIMPLE fixed report request-time reductions from few-column execution (C4) and packing of admitted groups (P4); the RTX C4 contrasts are descriptive secondary results. The fixed-crossing result demonstrates that attention can observe its physical-view boundaries while projection retains parallel work under the same numerical reference.
\end{abstract}
\maketitle
\hypersetup{pdfauthor={Chengzhan Li},pdftitle={Recovering Parallelism under Fixed Numerical References in Quantized LLM Decoders},pdfsubject={Named arXiv manuscript with scientific supplement},pdfkeywords={quantized inference, numerical reference, parallel execution, attention, KV cache, speculative decoding}}
\section{Introduction}
\label{sec:introduction}

Consider four known queries at consecutive positions. They can share one four-column projection group while preserving the reference arithmetic. Attention requires physical extents $[256,256,256,512]$. The first three queries use a 256-slot key--value (KV) view and the fourth uses a 512-slot view; the causal mask selects each query's visible tokens within that view. SIMPLE (the whole-call splitting baseline) divides the decoder into q3+q1 calls. This meets the attention requirement but also breaks the compatible projection group. The attention boundary thus restricts operations that could still execute together. Recovering that work requires separate partitions for projection and attention while preserving the scores and state returned to the consumer.

A fixed numerical reference requires execution to preserve the complete score rows requested by the consumer, committed active KV, and the consumer state passed to the next call. Two executions can select the same next token yet produce different scores or cache contents. Deterministic kernels and stateful inference systems already permit execution freedom under numerical constraints~\cite{thinkingmachines,llm42,vosti,detent}. In a quantized decoder, that freedom depends on the operator. Projections can share weight access and activation preparation when quantization, reduction and rounding match the reference. Attention also depends on physical extents, layouts and dispatch.

LOCAL retains the q4 projection, executes attention in two runs (three queries and one query), and joins their outputs in logical order (Figure~\ref{fig:construction}). Before native buffers are reused, it copies the requested score rows into owned storage for acceptance and rejection decisions. The consumer commits the accepted KV prefix and records any selected token whose KV is still to be computed. The next call uses that state and checks the complete extent vector before reusing its graph.

% Keep the contributions together after the introductory setup.
\par\penalty0\relax
\begin{samepage}
\begin{enumerate}
\item An admitted projection group can span several physical attention views. Counterexamples distinguish arithmetic mismatches from physical-view mismatches that remain after projection repair (Section~\ref{sec:reference}).
\item LOCAL preserves such a group across physical attention-view boundaries within one call. It splits only attention, rejoins outputs in query order, and returns owned score rows to the consumer, which commits the next-call state. The composition argument states sufficient operator and state conditions for preserving the fixed numerical reference interface across calls (Sections~\ref{sec:method}--\ref{sec:implementation}).
\item The fixed 32B crossing shows a supporting 21.56\% reduction in verification-through-commit time when LOCAL preserves q4 and SIMPLE uses q3+q1. The prespecified primary noncrossing control retains q4 in both arms, with an estimated 0.43\% reduction on the same clock. Across two fixed 8B conversations under target-only decoding, the primary equal-weight estimate is a 1.44\% reduction in response-plus-reset time (95\% CI [$-0.82\%$, $3.85\%$]). Separate request studies in Sections~\ref{sec:design}--\ref{sec:discussion} hold SIMPLE fixed to evaluate inherited few-column execution (C4) and packing of admitted groups (P4); the RTX C4 contrasts are descriptive secondary results.
\end{enumerate}
\end{samepage}

\section{Numerical Constraints on Arithmetic and Views}
\label{sec:reference}

\subsection{The interface to preserve}
The reference execution $R$ is fixed before transformation. Its profile specifies model and tokenizer bytes, weight quantization and scales, and activation preparation. It also specifies reduction and rounding behavior, the backend build and device, cache representation, and consumer rules. Within this numerical and state interface, execution may reorganize independent work and physical views. Graph structure, machine instructions, fusion decisions and launch sequences can therefore differ.

Let $Q=(x_0,\ldots,x_{q-1})$ be known tokens at contiguous positions, and $O$ the logical output rows requested by the consumer. In the evaluated interface, $O$ selects either all rows or the last row. Let $F$ denote the transformed execution. For complete binary32 vocabulary vectors $z_i$, canonical active KV $\mathcal K$, and consumer state $s$, the required relation at the corresponding output and commit boundaries is
\begin{equation}
 z_i^F\biteq z_i^R\;(i\in O),\qquad
 \mathcal K^F\biteq\mathcal K^R,\qquad s^F=s^R.
 \label{eq:contract}
\end{equation}
State $s$ records the \emph{materialized prefix}, whose K/V has been committed to the active cache, the prefix's position, and termination status. It also records a \emph{pending token}: a token selected for emission but not yet processed into that cache. Unused allocation is outside the active-state relation, while every operand actually exposed to attention, including physical padding, is governed by the profile. Complete-vector equality preserves scores for consumers beyond next-token selection.

{\looseness=-1
A conditional scalar reference checks each verification row under the same supplied proposal history, including prefixes later rejected (Equation~\eqref{eq:contract}). An autonomous target reference checks the emitted continuation and stopping behavior. The current CUDA construction assumes a single contiguous sequence, dense nonwrapping F16 KV, causal non-FlashAttention, and greedy consumers.\par
}

\subsection{Projection groups and attention views impose distinct constraints}
Admitting a projection requires its group width, quantization, scale application, partial reductions and epilogue behavior to satisfy the reference profile. In the captured historical CUDA panel of Figure~\ref{fig:constraints}(a), native widths two and four match concatenated scalar results in all 16 model/operator pairs per width. Native widths eight, sixteen, and thirty-two differ in every pair. Tiling or packing admitted four-column groups restores the recorded complete-operator outputs at the exercised wider widths. Independent admitted groups thus expose parallel work at wider request shapes.

\paperfigure[!t]{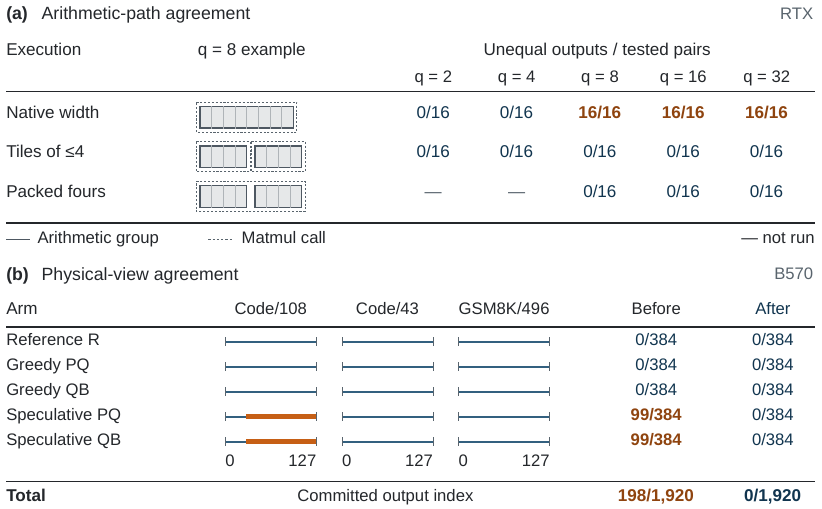}{Arithmetic and physical-view agreement. (a) Counts of complete operator outputs that differ from concatenated scalar columns, among 16 RTX model/operator pairs per width. Packed q2/q4 were not run. (b) Coordinated view repair removes 198 mismatches among 1,920 B570 committed vectors. Bold orange track segments locate the pre-repair mismatches; PQ/QB name the recorded attention layouts.}{The upper matrix compares native, tiled and packed projection arithmetic on RTX. Native q8, q16 and q32 each have 16 unequal complete outputs among 16 pairs; tiled and packed fours match, and packed q2/q4 are marked not run. The lower B570 tracks locate 99 mismatches at indices 29 through 127 on one of three inputs in each of two speculative arms. Counts total 198 of 1,920 vectors before physical-view repair and zero afterward. Bold orange segments mark pre-repair mismatches. The denominator includes three inputs, 128 rows and five arms including reference self-comparisons.}{fig:constraints}

\par\penalty0\relax
\begin{samepage}
Attention imposes a separate physical-view requirement. In the historical B570/Vulkan study, coordinated view repair restores agreement across all 1,920 complete committed vocabulary-vector comparisons (Figure~\ref{fig:constraints}(b)). Projection-compatible execution had left 198 mismatches, 99 in each of two arms. The repair changes extent/crop, layout, and dispatch together. A causal mask specifies logical visibility; the numerical contract also binds physical operands and the reduction schedule.\par
\end{samepage}

For the bound dense cache, a padding unit $P$, capacity $C$, and zero-based query position $t_i$ give the scalar append extent
\begin{equation}
 e_i=\min\!\left(C,\ P\left\lceil(t_i+1)/P\right\rceil\right).
 \label{eq:extent}
\end{equation}
The adapter obtains native requirements and validates this mapping. A physical-view descriptor also includes types, strides, head geometry, dispatch, and the KV values read. Adjacent queries with compatible descriptors form \emph{attention runs}; projection groups follow arithmetic admission. SIMPLE applies the attention partition to the entire decoder. LOCAL confines that partition to attention, then connects its outputs to the retained projection group and the next committed state.

\section{Local Attention Refinement and State Continuation}
\label{sec:method}

\subsection{From two partitions to one decoder transaction}
A decoder transaction returns the requested score rows in logical order and commits the active KV and consumer state for the next call. LOCAL keeps compatible queries in one outer call, splits their physical views within attention, and rejoins the attention outputs before the remaining layer operations. The consumer then uses the requested score rows to select and commit the continuation~(Algorithm~\ref{alg:lowering}).

\begin{algorithm}[!t]
\caption{Execute known queries under a registered reference profile}
\label{alg:lowering}
\small
\textbf{Inputs:} the profile specifying reference $R$, admitted recipes, incoming active KV and consumer state $s$, contiguous known rows $Q$, requested logical rows $O$, and fixed execution policy.
\begin{enumerate}
\item \textbf{Plan.} Validate cache, positions, capacity and output reserve; obtain each row's reference extent. Form outer calls of at most 32 rows: \Simple{} splits at incompatible physical requirements; \Local{} keeps the planned outer call intact across these boundaries.
\item \textbf{Execute each outer call.} Reuse its graph only if the complete extent vector and reference policy match and the shape and mask checks pass. In each layer:
\begin{enumerate}
\item Execute admitted projections and write current K/V.
\item Form adjacent compatible attention runs; slice Q/K/V and mask views, materialize noncontiguous masks, and execute admitted query tiles.
\item Join attention outputs in logical order before output projection and the remaining layer operations.
\end{enumerate}
After each call, synchronize and copy every requested complete vocabulary vector into its owned logical slot before the next native call.
\item \textbf{Commit.} Apply the registered consumer to the owned rows. Trim tentative KV, retain the accepted materialized prefix, and record pending token and termination. Return owned outputs and committed state to the successor's planning step.
\end{enumerate}
\end{algorithm}

\emph{Owned score rows} are complete requested vocabulary vectors copied from reusable native storage into logical output slots. These copies remain available for acceptance, correction or bonus selection after another native call reuses that storage. Commit records the materialized prefix and pending token defined in Section~\ref{sec:reference}.

In the fixed A5000 crossing case, a pending seed and frozen proposals $p_1,p_2,p_3$ occupy positions 253--256, with extents $[256,256,256,512]$. \Simple{} executes a three-row decoder call followed by a one-row call. \Local{} executes one four-row call with two attention runs (three queries and one query), retaining the admitted four-column projection (Figure~\ref{fig:construction}).

\paperfigure{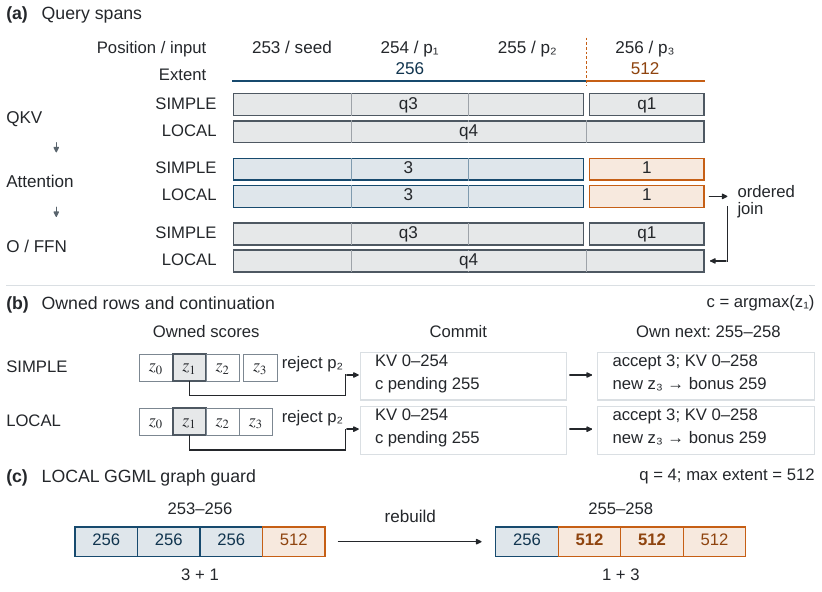}{LOCAL at the fixed A5000 32B crossing (C4 on, P4 off). (a) The seed and proposals $p_1,p_2,p_3$ share query coordinates across SIMPLE's q3+q1 calls and LOCAL's q4 projection with $3+1$ attention views. (b) Within each transaction, $z_i$ scores the token after zero-based input row $i$: the consumer uses $z_0$ to accept $p_1$ and $z_1$ to reject $p_2$. SIMPLE's owned copy of $z_1$ supplies the scores for correction $c$ after native buffer reuse. Each arm's successor accepts three new proposals. (c) LOCAL rebuilds its views when the extent vector changes. Spans encode query membership.}{Three panels show the crossing construction, a rejected proposal and the successor guard. A pending seed and proposals p1, p2 and p3 occupy positions 253 to 256. SIMPLE makes q3 and q1 decoder calls while LOCAL makes a q4 call with 3+1 attention views and rejoins before output projection and FFN. Score indices start at zero within each transaction; zi predicts the token following input row i. SIMPLE preserves its first call's owned score row z1 by synchronization and copying. The consumer accepts p1 using z0, then rejects p2 using z1, leaving KV positions 0 to 254 and correction c pending at 255. Each arm continues from its own commit at positions 255 to 258 and accepts three new proposals. LOCAL rebuilds 1+3 views after the full extent vector changes. The successor's z3 selects the bonus pending at 259. Box spans encode query membership, not time or memory size; the two policy arms are alternatives.}{fig:construction}

\subsection{Projection groups across attention boundaries}
Each layer computes its admitted projections and writes current keys and values to the canonical cache before attention reads them. LOCAL forms adjacent compatible attention runs from the complete extent vector and bound view descriptors. It slices Q, K, V and the mask for each run, materializes a contiguous mask where required, and tiles the run at the admitted width (at most four queries in the exercised layout). Within each tile, queries occupy an outer tensor dimension and retain their one-column attention matrix operations.

For a selected row, the causal mask marks future and empty cache slots with $-\infty$. On the evaluated non-FlashAttention path, QK uses the bound dot-product recipe. When QK scores are finite and a visible key exists, the scaled masked softmax assigns the masked positions zero weight. PV then multiplies these weights by the bound value view; finite values in masked slots contribute zero-weight products. The same reference recipe binds extents, strides and reduction order. The released \href{https://github.com/leecdiang/RefBound/blob/58697de93d86fc39c105bdfb8d8c0fd2949f192c/runtime/backend/src/llama-kv-cache.cpp\#L1594-L1698}{causal mask} and \href{https://github.com/leecdiang/RefBound/blob/58697de93d86fc39c105bdfb8d8c0fd2949f192c/runtime/backend/src/llama-graph.cpp\#L2820-L2866}{non-FlashAttention operations} specify this recipe; qualification checks complete output rows and active state.

Ordered assembly concatenates the attention results in logical query order before output projection and feed-forward computation. Normalization, residual connections and inherited layer operations follow the reference profile. \Simple{} splits the whole decoder at each incompatible view boundary and starts a new call there. Both policies request identical logical rows. The q4 versus q3+q1 comparison changes projection widths, dispatch, fusion and launches together, and includes view construction, copies, joins and synchronization.

\subsection{Score-row ownership across rejection and trim}
The frontend synchronizes after each native call and copies every requested complete vocabulary vector into an owned logical slot. These copies survive native storage reuse. Under \Simple{}, the first subcall's rows remain available after the second executes.

Rows $z_0,\ldots,z_3$ use zero-based indices within each transaction and score the token following their input row. In Figure~\ref{fig:construction}(b), the consumer accepts $p_1$ using $z_0$, then rejects $p_2$ and selects correction $c$ using $z_1$. SIMPLE produced $z_1$ in its three-row subcall. Its final one-row subcall returns $z_3$ under a history that includes $p_2$ and $p_3$. The owned copy preserves the row needed for that decision. Trimming retains KV at positions 0--254 and leaves the correction pending at 255. Each arm starts its successor from its own retained KV and pending correction. That successor processes positions 255--258. After accepting all three new proposals, the consumer selects the bonus from row 3 and leaves it pending at 259.

\par\penalty0\relax
\begin{samepage}
\subsection{Graph reuse under the successor's views}
With extent vector $[256,512,512,512]$, the successor needs $1+3$ attention views, although its query count, maximum extent and reference policy match the preceding transaction. LOCAL compares the complete extent vectors, rejects the preceding decoder graph and builds the successor's $1+3$ partition in place of $3+1$.\par
\end{samepage}

The reuse predicate checks \texttt{prefix\_n\_kv}, \texttt{reference\_policy}, and inherited shape and mask conditions. The choice of LOCAL or SIMPLE and the packing policy stay fixed within a worker, with context modes set at decode boundaries. The resulting graph reads the active KV supplied by that arm's commit.

\subsection{Composition from operator outputs to successor state}
\paragraph{Sufficient conditions.}
Fix a reference execution $R$, its profile and an input domain. Suppose that (i) candidate and reference process the same known rows $Q$ and requested rows $O$, entering with equal active KV and consumer state; (ii) every admitted operator preserves $R$'s outputs and KV writes on the inputs it receives; (iii) the row partitions are complete, disjoint and reassembled in logical order; (iv) requested rows remain owned until consumed; and (v) both executions apply the same consumer transition. Then the composed transaction preserves Eq.~\eqref{eq:contract}, and its successor starts from the same committed state.

Induction over the decoder graph establishes equality at each operator boundary. Ordered reassembly and ownership preserve the consumer's row identity. Equal complete vectors produce equal greedy decisions. The common trim, pending-token and stopping rules then establish the next induction base. Operator equivalence is assumed on the admitted input domain. This composition follows a stateful-contract perspective~\cite{vosti}; LOCAL supplies an executable construction across nonaligned projection and attention partitions. Table~\ref{tab:obligations} maps these operator and state obligations to their implementation and checks.

% Keep this table beside its first callout with compact, local float spacing.
\begingroup
\setlength{\intextsep}{3pt plus 1pt minus 1pt}
\begin{table}[H]\caption{Operator and state obligations, their implementation, and the checks that exercise them.}\label{tab:obligations}% Native table source; current deployment identities: evidence/qualification_manifest.json.
\begingroup
\setlength{\tabcolsep}{3pt}\renewcommand{\arraystretch}{1.05}
\begin{tabularx}{\linewidth}{@{}P{.18\linewidth}P{.40\linewidth}Y@{}}
\toprule
Obligation & Implemented connection & Executed check and scope\\
\midrule
Admitted arithmetic & Tensor predicates; inherited Q8\_1 preparation, scales, reductions and epilogues; disjoint P4 groups & Complete operator and vocabulary comparisons at qualified widths; RTX C4 and A5000 P4 device dispatch\\
Physical attention view & Native extent query; adjacent-run partition; mask slicing; ordered assembly & Fixed-crossing R/SIMPLE/LOCAL complete output rows and active KV match byte for byte\\
Logical row ownership & \texttt{PhaseContext::decode} synchronizes, then copies each full row before another call & Full-row comparisons and recomputed rejection, correction and bonus selection\\
Commit and pending token & \texttt{real\_sd::select} chooses keep/pending; each arm trims its own cache & Complete committed KV and pending state match the independent reference\\
Own successor and reuse & Each arm carries its own keep/pending; graph guard binds the complete extent vector & $3+1\rightarrow1+3$ crossing replay and retained-prefix comparisons; each arm continues from its own commit\\
\bottomrule
\end{tabularx}
\par\smallskip\noindent
Section~\ref{supp:access} binds these checks to their tested inputs, binary identities and complete-object comparisons.
\endgroup
\end{table}
\endgroup

\section{Implementation and Numerical Qualification}
\label{sec:implementation}

\subsection{Arithmetic paths after shared preparation}
A llama.cpp-derived decoder uses existing quantized kernels to implement LOCAL's distinct projection and attention partitions and state continuation~\cite{llama_cpp_fewrow,loop4code}. \Cfour{} selects admitted few-column paths for two to four queries; \Pfour{} organizes independent four-column groups in wider prefill; \Local{} keeps projection groups intact across attention-view boundaries.

The RTX controls share Q8\_1 activation preparation (Figure~\ref{fig:c4}a). For $q$ queries, the backend scalar-loop control B invokes the existing scalar specialization $q$ times. C1 executes $q$ independent scalar query/channel groups in one dispatch, indexed by \texttt{grid.y}. C4 invokes the inherited few-column specialization once. Initial and confirmation RTX studies use the same numerical core, worker binaries and models. Target-only (\TO{}) and speculative (\SD{}) consumers are chosen independently of these projection paths.

The admission predicates bind the arithmetic path to a tensor and device configuration. Here $K$ is the reduction dimension, $M$ the output dimension, and $q$ the number of query columns:
\begin{center}
\begin{tabularx}{\linewidth}{@{}P{.18\linewidth}YY@{}}
\toprule
Condition & A5000 C4 & RTX C4\\
\midrule
Weights & Q4\_K or Q6\_K & Q8\_0\\
Reduction & $K\geq2048$, $K\bmod256=0$ & $K\geq1024$, $K\bmod32=0$\\
Output/width & $M$ even; $2\leq q\leq4$ & $M$ even; $2\leq q\leq4$\\
Layout & All tensors contiguous; non-view weights; F32 input/output & Contiguous input; single weight/input channel and sample\\
Dispatch & Marked, unfused matmul; no IDs & Marked, unfused matmul; no IDs\\
\bottomrule
\end{tabularx}
\end{center}
Both paths require the GENERIC device table and F32 input/output; shape predicates require matching extents. On A5000, C1 retains the scalar-channel representation, and P4 is disabled in both the C1 and C4 arms. RTX also disables FlashAttention, CUDA graphs and programmatic dependent launch. The inherited few-column path is admitted under these conditions. The source identities and exercised cases are recorded in Section~\ref{supp:access}.

\subsection{Packing independent groups with admitted arithmetic}
For $q=4g$ known columns, P4 interprets supported contiguous activation columns as $g$ independent four-column groups after shared preparation. Each group retains its admitted arithmetic, with no reduction across groups. The A5000 graph keeps input and output shapes $[K,1,q]$ and $[M,1,q]$. At launch, the quantized matrix--vector multiplication (MMVQ) dispatcher interprets these columns as groups. The wide predicate is $8\leq q\leq32$, \mbox{$q\bmod4=0$}; other shapes retain their registered paths. Table~\ref{tab:qualification-details} records exercised widths and complete-object qualifications.

The LOCAL study uses C4 with P4 disabled and reuses the frozen A5000 prefill numerical core. At its crossing, SIMPLE uses the few-column path for q3: it records 449 q3 and 385 q1 projection dispatches, versus LOCAL's 449 q4 dispatches. SIMPLE's q1 fragment partly uses gate/up fusion, while LOCAL's q4 path is unfused. The separate RTX B/C1/C4 controls disable fusion.

Both crossing policies execute two logical attention runs per layer. LOCAL places them inside one graph, with four refinement views per tile and one ordered join per layer. Its initial $3+1$ split also materializes the noncontiguous three-row mask; its successor's $1+3$ split needs no such mask copy. These source-derived counts describe refinement nodes in addition to inherited attention operations. The transaction clock includes their execution, synchronization and commit.

\subsection{Device dispatch and state agreement}
Qualification compares complete requested vocabulary vectors, active KV in canonical logical order, and each arm's continuation from its own commit. Captures follow the conditional and autonomous references defined in Section~\ref{sec:reference}.

On RTX 5070 Ti, CUPTI traces for the fixed 757-token input identify 8,550 q1 Q8\_0 MMVQ verification launches in C1, and 8,325 q4 plus 225 q2 launches in C4. The 75 corresponding intermediate active F16 KV objects per arm match byte for byte across C1 and C4. Their canonical hashes also match the historical hashes. Final state matches the independent reference, and the continuation checks pass. In each arm, 174 complete F32 target score vectors match the independent q1 replay bit for bit.

On A5000, the fixed crossing exercises the planned reference ($R$), SIMPLE and LOCAL paths. All 18 complete F32-vector pairs and ten active-KV pairs match the independent q1 reference. Each arm continues correctly from its own committed state.

The unmodified Linux/sm86 RefBound P4 candidate passes a seven-arm qualification. The P4-enabled target-only and live-SD runs each record 1,780 P4 dispatches in q32 target prefill. The 2,125 byte-for-byte comparisons cover full F32 vectors, active KV, independent q1 controls and continuation. The q31 control exercises the registered fallback and checks its own state contract; independent numerical qualification of the release ELF covers q32. Section~\ref{supp:access} gives the objects, comparison scopes and binary identities.

\section{Evaluation Design}
\label{sec:design}

The LOCAL studies compare fragmented and retained projection work at fixed histories, then count excess verification calls and measure native-service time in M1 and response-plus-reset time in the natural 8B study (Table~\ref{tab:study-map}). The controlled 32B LOCAL comparisons use C4 enabled and P4 disabled. Separate C4 and P4 studies hold SIMPLE fixed to measure request-time changes from arithmetic and packing. Each study uses its own contemporaneous control. Tables~\ref{tab:deployment-details} and~\ref{tab:study-details} give the model, placement, workload and sampling inventories.

\begin{table}[tbp]\caption{Study questions, controls, timed endpoints and statistical roles. TO is target-only; SD is speculative decoding.}\label{tab:study-map}% Author-edited study map; configuration and value records remain in supplement tables.
\begingroup
\fontsize{8.7}{10.5}\selectfont
\setlength{\tabcolsep}{3pt}\renewcommand{\arraystretch}{1.04}
\begin{tabularx}{\linewidth}{@{}P{.17\linewidth}P{.14\linewidth}P{.19\linewidth}P{.14\linewidth}Y@{}}
\toprule
Question / study & Change & Hold fixed & Endpoint & Units and statistical role\\
\midrule
Direct LOCAL, A5000 32B & SIMPLE / LOCAL & C4 on; P4 off; frozen proposals & Verify through commit & Two histories, 12 paired blocks; noncrossing primary, crossing and interaction supporting. Original two-repeat pairs descriptive.\\
\addlinespace[4pt]
Natural LOCAL, A5000 8B & SIMPLE / LOCAL & Target-only; two fixed conversations & Response plus reset & Two-turn episodes, 12 blocks; equal-conversation primary, each conversation supporting.\\
\addlinespace[4pt]
Natural LOCAL, A5000 M1 & WHOLE / LOCAL $\times$ loop4 / packed & Q8 profile; same roots and sessions & Native service & 16 roots $\times$ 4 sessions; five-comparison family, 99\% marginal intervals.\\
\midrule
Few-column value, RTX confirmation & B / C1 / C4 & Shared preparation; SIMPLE; P4 off & Resident request & 24 roots $\times$ 6 blocks $\times$ 2 rounds; C1/B primary, C4 contrasts descriptive.\\
\addlinespace[4pt]
Earlier C4 requests & C1 / C4; RTX also B & SIMPLE; P4 off & Resident request & A5000 12 roots, RTX 24; two rounds each, separate from confirmation.\\
\addlinespace[4pt]
Packing value, A5000 / RTX & TO / SD $\times$ C4 / P4 & SIMPLE; same consumer within each packing contrast & Resident request & 12 roots $\times$ 4 rounds (A5000); 24 $\times$ 2 (RTX); paired contrasts within each platform.\\
\midrule
Complete cost, B570 & TO / SD $\times$ A / C & Formal: Vulkan target; CPU draft & Compute plus teardown & Formal: 24 roots $\times$ 2 rounds. Separate draft-placement pilot: 2 roots $\times$ 1, stopped by service criterion.\\
\bottomrule
\end{tabularx}
\par\endgroup
\end{table}

\subsection{Controls and workloads}
Modern A5000 studies use Qwen3-32B Q4\_K\_M with Q8\_1 activation preparation and a Qwen3-0.6B Q8\_0 draft~\cite{qwen3}. RTX uses Llama-3.1-8B-Instruct and Llama-3.2-1B-Instruct, both Q8\_0~\cite{llama3,llama32}. Their C4/P4 panels use fixed, previously exposed workload inputs (roots) from CodeContests and MATH~\cite{codecontests,math}. The LOCAL comparison freezes proposals from a real speculative trace at histories 253 and 137 of one root; successors follow their own arm's commit. The histories differ in content as well as crossing geometry, so their interaction applies to these two fixed cases.

A5000 confirmation A repeats the two 32B transactions over 12 paired time blocks. Its frozen plan prioritized the uncertain noncrossing effect in the original two-repeat study. Noncrossing is the primary endpoint, while crossing and their interaction are prespecified mechanism-supporting endpoints. Confirmation B retains the original 8B Q8\_0 target-only consumer and evaluates two fixed OpenAssistant conversations~\cite{oasst}. Each conversation has two turns, a context capacity of 4,096 tokens (C4096), and a cap of 128 output tokens per turn. Its 12-block primary endpoint weights the conversations equally. The separate RTX confirmation uses 24 roots, six blocks and two rounds for each root and block, giving 864 measured requests. C1/B is its sole prespecified primary comparison. C4/C1 and C4/B are descriptive secondary comparisons.

\par\penalty0\relax
\begin{samepage}
The historical A5000 study M1 crosses WHOLE/LOCAL and loop4/packed implementations on 16 APPS/MATH roots with four sessions each~\cite{apps,math}. WHOLE is its name for the whole-call splitting strategy used by SIMPLE. Its Qwen3-8B/0.6B Q8\_0 profile differs from the modern CUDA studies. B570 uses its own Vulkan F32-by-Q8 reference and TO/SD $\times$ A/C design, with scalar target A, compatible grouped target C and a CPU draft. The later two-root pilot changes draft placement.\par
\end{samepage}

\subsection{Measurement boundaries}
CUDA resident-request time runs from reset through prefill, target evaluation, live drafting where present, requested-output handling, decisions, commit and final synchronization. Loading, context-pool setup and post-clock exports are outside it. In the controlled LOCAL/SIMPLE transaction study, the clock runs from target verification through trim and synchronization, including owned-row copies and decisions; history reconstruction and live drafting are excluded. Profiler diagnostics from the original two-repeat transaction study use a separate device-event clock.

\begin{samepage}
M1 uses native-service time. The natural 8B study measures response plus reset, including shared greedy-sampling and control work. B570 service includes CPU drafting, target compute and normal teardown, excluding intervening endpoint exports. Worker totals also include loading, checks and preparatory primer calls.\par
\end{samepage}

\subsection{Estimates and dependence}
A \emph{root} is a workload input, a \emph{history} a fixed prefix, a \emph{round} a repetition of matched arms, and a \emph{time block} a scheduled group of repeats. A two-turn \emph{episode} is the natural-confirmation unit. Repeating a fixed history across blocks measures temporal variation without adding workload inputs.

For paired candidate/control times at root $i$ and round $j$, modern request estimates give roots equal log weight:
\begin{equation}
 \ell_{ij}=\log\frac{T^{\mathrm{candidate}}_{ij}}{T^{\mathrm{control}}_{ij}},\qquad
 \widehat r=\exp\!\left(\frac1n\sum_i\frac1{m_i}\sum_j\ell_{ij}\right).
 \label{eq:estimator}
\end{equation}
Time reduction is $100(1-\widehat r)\%$. Initial request intervals resample whole roots within code/math strata, retaining rounds. RTX confirmation averages its two rounds within each root/block, then weights 24 roots and six blocks equally; its crossed bootstrap independently resamples stratified roots and whole blocks. Root-only intervals condition on the observed blocks. A5000 confirmation resamples its 12 blocks jointly, preserving both histories or conversations. These individual 95\% intervals describe the fixed cases in the observed campaign. Arithmetic phase differences and geometric time ratios are separate estimands.

For M1, all four sessions remain within their root. Direct effects use ratios of total native-service time. For each root, the interaction divides the LOCAL/WHOLE time ratio under packing by the corresponding ratio under loop4. The reported interaction is the geometric mean of those quotients. Its five-comparison Bonferroni family uses 99\% marginal intervals. The supplement gives formulas, seeds and draws; original two-repeat LOCAL results remain descriptive pairs.

Qualification, warmup and diagnostics are excluded from timing estimates. Timing estimates use every eligible measured root and block; the ledger records technical replacements and scientific stops (Section~\ref{supp:ledger}). M1's released request records give call counts but omit the cycle-level partitions and timings needed to estimate the time share of fragmented work. Section~\ref{supp:access} identifies which intervals are reconstructed from released scalars and which retain frozen historical endpoints.

\section{Results}
\label{sec:results}

\subsection{Local refinement retains q4 work at a physical crossing}
\label{sec:local}
At the fixed 32B crossing, LOCAL retains the q4 projection work that SIMPLE splits into q3+q1. Its supporting verification-through-commit endpoint shows a 21.56\% reduction [20.21\%, 22.83\%] over 12 paired A5000 time blocks (Figure~\ref{fig:confirmation}). The primary noncrossing history retains q4 under both policies, providing a comparison without projection-group fragmentation. Its estimated reduction is 0.43\%, with a 95\% interval of $[-1.13\%,2.01\%]$. For these two fixed histories, the supporting interaction divides the crossing's geometric-mean LOCAL/SIMPLE time ratio by the noncrossing ratio: 0.788 (95\% CI [0.768, 0.806]). Values below one mean a larger relative reduction at the crossing.

\paperfigure[!htbp]{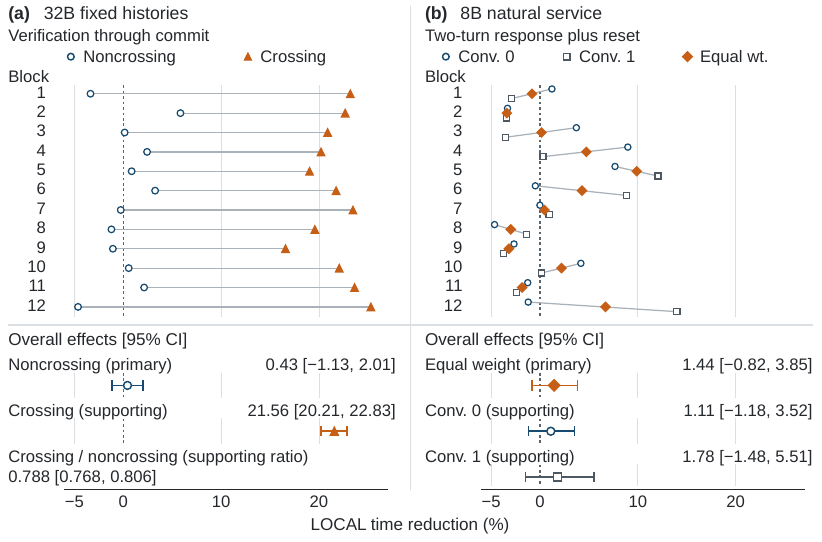}{LOCAL/SIMPLE time reductions over 12 paired A5000 blocks: (a) fixed 32B transactions through verification and commit; (b) natural 8B target-only episodes through response and reset. Lines pair cases within each block. Positive values favor LOCAL; printed limits are individual 95\% whole-block intervals. Primary and supporting roles are marked.}{Two panels show all twelve numbered time blocks. Lines connect cases within each block. The 32B crossing effects consistently favor LOCAL, while the primary noncrossing interval spans zero. The separately clocked 8B target-only panel shows both conversations and their equal-weight diamond; its primary interval also spans zero. Printed intervals accompany the estimates, and the supporting ratio-of-ratios interaction is below one.}{fig:confirmation}

Device-event diagnostics from the original two-repeat study show where the time changes occur. At the crossing, projection time falls from 74.334 to 53.417\,ms as q3+q1 becomes q4; attention time changes from 12.060 to 11.621\,ms, with 0.134\,ms of added concatenations. These component timings measure the lowering as a whole, including its changes in widths, fusion, launches and views. Section~\ref{supp:local} reports the original paired transaction times and the separately clocked device diagnostics.

The successor to each first transaction continues from that transaction's committed state, exercising score ownership, trim and the view guard. Table~\ref{tab:structure-cost} relates the retained work to its cost and frequency in each deployment.

\begin{table}[tbp]\caption{Retained work, execution cost and observed outcome under each profile. Counts and timings use the study units shown.}\label{tab:structure-cost}% Native table source. Modern traces: analysis/n01/results/STRUCTURE.json.
\begingroup
\setlength{\tabcolsep}{3pt}\renewcommand{\arraystretch}{1.05}
\begin{tabularx}{\linewidth}{@{}P{.16\linewidth}P{.29\linewidth}YY@{}}
\toprule
Profile / case & Preserved work & Cost and opportunity & Observed result\\
\midrule
A5000 32B, fixed crossing & q3+q1 $\rightarrow$ q4; both q3 and q4 use admitted few-column arithmetic; 834 $\rightarrow$ 449 projection dispatches & Two attention runs/layer in either arm; LOCAL adds internal views, mask materialization and joins & Supporting transaction reduction 21.56\% [20.21\%, 22.83\%]\\
A5000 32B, noncrossing & q4 $\rightarrow$ q4; 449 dispatches in either arm & One intact q4 group; one attention run/layer & Primary transaction reduction 0.43\% [$-1.13$\%, 2.01\%]\\
A5000 8B, natural & Outer calls: Conv.\ 0, 263 $\rightarrow$ 263; Conv.\ 1, 273 $\rightarrow$ 272 & Decode, sampling, controller and reset remain inside service & Primary service reduction 1.44\% [$-0.82$\%, 3.85\%]\\
M1, natural & 2,244 $\rightarrow$ 2,224 calls for 2,224 logical cycles per arm & 20 excess calls per 2,224 cycles; request-level cost records & LOCAL service reductions 0.209\% (loop4), 0.128\% (packed)\\
Historical Vulkan, q32 crossing & Recorded q32 geometries include $16+16$ whole-call splits & Recorded per-cell transaction times & LOCAL/WHOLE ratios 1.055, 1.015, 1.089 for the three recorded cells\\
\bottomrule
\end{tabularx}
\par\smallskip\noindent
Modern counts come from qualification traces and the original profiler; node counts are source-derived. Printed intervals are individual 95\% block intervals (Figure~\ref{fig:confirmation}). M1 records omit cycle partitions and timings; historical Vulkan records omit per-cell dispatch and view/copy/join costs. Tables~\ref{tab:m1} and~\ref{tab:historical-geometry} give their timings and intervals. Each profile retains its own model, geometry and clock.
\endgroup
\end{table}

\subsection{Natural opportunities and service costs}
\label{sec:natural}
In M1, LOCAL removes 20 extra physical verification calls across 2,224 logical cycles. The excess-call count is 0.90\% of the logical-cycle count. Each arm of the WHOLE/LOCAL $\times$ loop4/packed factorial has the same 2,224 cycles (Table~\ref{tab:m1}). WHOLE executes 2,244 physical verification calls and LOCAL 2,224.

\begin{table}[!htbp]\caption{M1 factorial: 16 roots and four sessions per root, with 64 measured requests per cell. Each projection row reports native-service totals and its direct LOCAL/WHOLE time reduction. Intervals retain the frozen historical 99\% endpoints of the five-comparison Bonferroni family. Table~\ref{tab:m1-family} gives all five comparisons.}\label{tab:m1}% Table 5 | tab:m1 | editable body generated by scripts/build_tables.py
\begingroup
\fontsize{8.7}{10.5}\selectfont
\setlength{\tabcolsep}{3.5pt}\renewcommand{\arraystretch}{1.03}
\noindent
\begin{tabularx}{\linewidth}{@{}>{\raggedright\arraybackslash}Xrrrrr@{}}
\toprule
Projection & \multicolumn{2}{c}{Native service total (s)} & \multicolumn{3}{c}{LOCAL time reduction (\%)}\\
\cmidrule(lr){2-3}\cmidrule(lr){4-6}
organization & WHOLE & LOCAL & Estimate & 99\% low & 99\% high\\
\midrule
loop4 & $142.271$ & $141.974$ & $0.209$ & $0.042$ & $0.463$\\
packed & $139.945$ & $139.765$ & $0.128$ & $-0.057$ & $0.405$\\
\midrule
\multicolumn{6}{@{}l}{\textbf{Verification calls removed:} $20$ per $2{,}224$ logical cycles (0.90\%) per arm.}\\
\multicolumn{6}{@{}l}{Physical calls: WHOLE $2{,}244$, LOCAL $2{,}224$.}\\
\bottomrule
\end{tabularx}
\par\smallskip
\fontsize{8}{9.6}\selectfont\raggedright
\textit{Estimator.} Reductions use $100(1-\mathrm{LOCAL/WHOLE})$ with sums over the same roots and sessions. Historical loop4/packed name the M1 projection organizations. The interaction uses a separate root-geometric estimator.
\par\endgroup
\end{table}

At fixed loop4, LOCAL reduces total native-service time by 0.209\%, with 99\% interval [0.042\%, 0.463\%]. At fixed packed projection, the reduction is 0.128\%, with interval $[-0.057\%,0.405\%]$. Packing improves both cut policies. The root-geometric interaction is 1.00070 [1.00007, 1.00139], a 0.070\% ratio change toward a smaller LOCAL effect under packing.

The 8B target-only confirmation measures two fixed conversations over 12 blocks. Its primary equal-conversation reduction is 1.44\%, with 95\% interval $[-0.82\%,3.85\%]$ (Figure~\ref{fig:confirmation}b). Conversation 0 uses 263 calls under either policy. For the same 809 target rows in Conversation 1, SIMPLE makes 273 calls and LOCAL 272. Differences in the shared greedy-sampling stage are comparable to the net service differences, even in Conversation 0, where grouping is unchanged (Table~\ref{tab:confirmation-costs}). The service estimates include variation in work shared by both policies. Section~\ref{supp:natural} and Tables~\ref{tab:confirmation-pairs}, \ref{tab:confirmation-costs} and~\ref{tab:natural} retain complete phase accounts, paired observations and the earlier natural panels.

% Keep the RTX heading and its complete opening comparison below earlier floats.
\par\penalty0\relax
\begin{samepage}
\subsection{Few-column execution versus scalar regrouping on RTX}
\label{sec:c4}
The RTX request study compares few-column C4 with the scalar loop B and grouped scalar C1, using shared activation preparation with SIMPLE fixed and P4 off. C4/C1 and C4/B are descriptive secondary contrasts. Their resident-request time reductions are 15.80\% [13.07\%, 18.30\%] and 15.04\% [12.40\%, 17.49\%], respectively. All 24 root aggregates favor C4 against either scalar control (Figure~\ref{fig:c4}).\par
\end{samepage}

The sole prespecified primary comparison, C1/B, estimates a 0.90\% request-time increase, with a 95\% interval from a 0.48\% reduction to a 2.19\% increase. Relative to B, C1 saves 100.43\,ms in target verification on average; target prefill adds 173.57\,ms and drafting adds 7.43\,ms. The arithmetic mean differences across all phases sum to an 80.52\,ms service increase (Figure~\ref{fig:c4}c).

% Finish earlier evidence floats before the C4 figure.
\FloatBarrier
\paperfigure[!htbp]{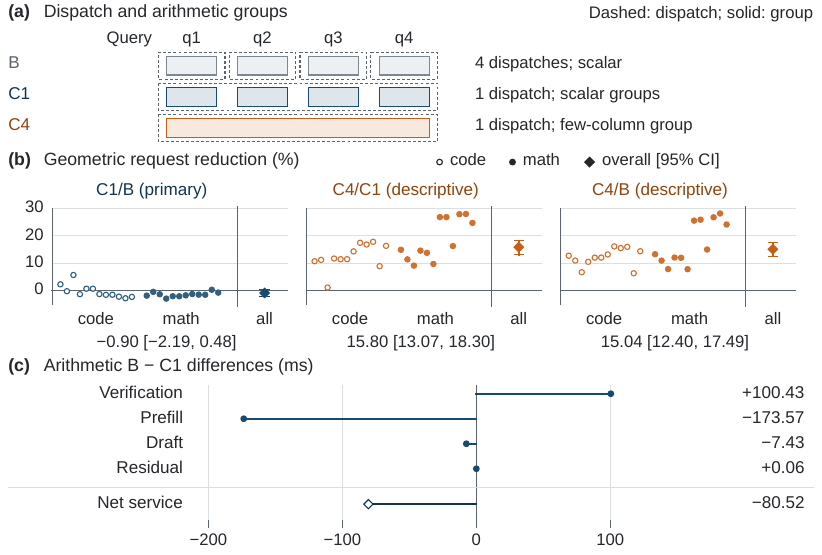}{RTX resident-request controls, SIMPLE fixed and P4 off. (a) Shared q1--q4 columns align dashed dispatch boundaries and solid arithmetic groups. (b) Root effects, geometric estimates and individual 95\% crossed root/block intervals, with statistical roles marked. Open/filled circles denote code/math roots. (c) Arithmetic mean B-minus-C1 phase differences from zero. Positive values favor C1. The net $-80.52$\,ms diamond denotes an 80.52\,ms increase in C1 service time.}{The dispatch diagram distinguishes scalar loop B, grouped scalar C1 and few-column C4 under shared q1 to q4 column headers. Each contrast has twelve code and twelve math root effects and a geometric summary. Open and filled circles denote code and math roots, respectively. The primary C1/B interval crosses zero; both descriptive C4 intervals favor the candidate. Independent zero-based phase rows show verification savings of 100.43 milliseconds, prefill minus 173.57, draft minus 7.43, residual plus 0.06 and net service minus 80.52. The open diamond at minus 80.52 milliseconds means that C1 service takes 80.52 milliseconds longer than B. These phase differences compare B with C1.}{fig:c4}

C1/B lowers verification time in all 24 root aggregates and all six block aggregates, but lowers service time in five roots and one block. C4/C1 reduces both verification and service time in every root and block (Table~\ref{tab:rtx269-phases}). Root and block aggregates summarize the same requests along two axes. C4 is used during verification and, for one root, the q3 tail of prefill.

\begin{samepage}
In the separate A5000 C1/C4 study with SIMPLE fixed and P4 off, C4 reduces request time by 27.77\% [25.43\%, 30.09\%] across 12 roots and two rounds; all few-column admissions occur in verification. Figure~\ref{fig:c4-initial} and Table~\ref{tab:direct-effects} report that study and the initial two-round RTX results.\par
\end{samepage}

% Keep the complete P4 setup together under the normally spaced heading.
\subsection{Packing within decoding policies and speculation with P4 fixed}
\label{sec:p4}
\begingroup
\edef\layoutparindent{\the\parindent}
\noindent\begin{minipage}{\linewidth}
\setlength{\parindent}{\layoutparindent}
\noindent P4 extends this organization to wide prefill by packing independent four-column groups while retaining each group's arithmetic.

The four-arm request studies first compare packed and unpacked execution within target-only and speculative decoding, then compare the two decoding policies with packing enabled (Figure~\ref{fig:p4}). All three paired contrasts use 12 A5000 roots and 24 RTX roots, with SIMPLE fixed.
\end{minipage}\par
\endgroup

\paperfigure[!htbp]{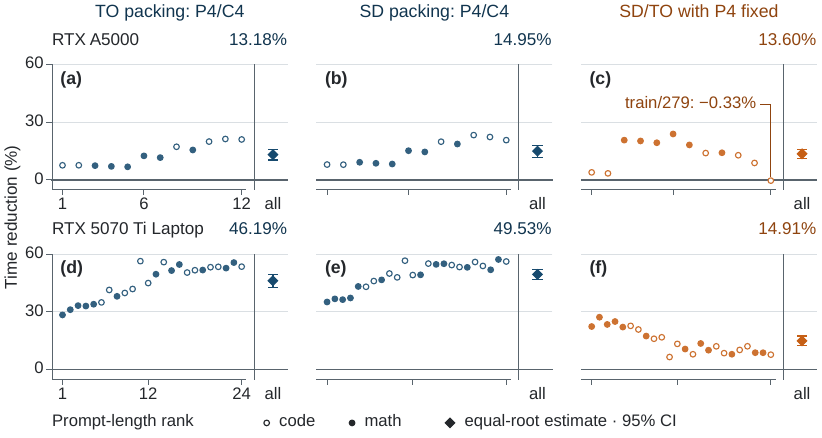}{P4 complete-request effects with SIMPLE fixed. Each platform row shares its own ordinal prompt-length ranks across three contrasts. Open/filled circles denote code/math roots; diamonds and bars give equal-root estimates and 95\% stratified-root intervals. Printed percentages label the six estimates; Table~\ref{tab:direct-effects} gives their intervals.}{A5000 occupies the top row and RTX the bottom row. The first two columns compare packing within TO and SD; the third compares SD/TO with P4 fixed. The 108 root-effect marks reuse 12 A5000 and 24 RTX roots in three comparisons. Open and filled circles denote code and math; diamonds and six printed percentages denote equal-root estimates. Prompt-length rank is ordinal within each platform and shared across its three contrasts. All panels share the time-reduction scale. Packing favors every root in the first two columns. In panel c, the open-circle code root train/279 at rank 12 is annotated minus 0.33 percent. Each platform has its own paired controls.}{fig:p4}

A5000 P4 reduces target-only request time by 13.18\% [10.38\%, 15.73\%] and speculative request time by 14.95\% [11.80\%, 17.76\%] relative to C4. The corresponding RTX reductions are 46.19\% [42.72\%, 49.32\%] and 49.53\% [47.00\%, 51.87\%]. Each contrast uses the contemporaneous unpacked control in the same decoding policy.

\begingroup
With P4 enabled in both arms, SD+P4 reduces complete-request time relative to TO+P4 by 13.60\% on A5000 [11.26\%, 15.89\%] and 14.91\% on RTX [12.50\%, 17.44\%]. These estimates include drafting and verification. Eleven of the twelve A5000 roots are faster; \texttt{train/279} is 0.33\% slower.\looseness=-1\par
\endgroup

On A5000, packing reduces speculative-request time by an arithmetic mean of 1,077.1\,ms; prefill accounts for 1,064.5\,ms of that reduction. All 232 prefill calls that invoke P4 use aligned q32 groups within one physical extent. RTX speculative prefill falls from 6.852 to 2.395\,s. Section~\ref{supp:effects} gives the complete phase accounts and observed dispatch widths.

% Complete the P4 evidence before the B570 service comparison.
\FloatBarrier
\subsection{Complete service favors target-only execution on B570}
\label{sec:cost}
\begin{samepage}
Across 24 roots and two rounds, compatible grouping reduces target-only time by 21.73\% [20.35\%, 23.01\%] and speculative time by 12.78\% [12.56\%, 12.99\%]. Including CPU drafting and teardown, speculative service takes 2.512 times grouped target-only time \mbox{[2.486, 2.539]}; all 24 roots are slower (Figure~\ref{fig:b570}).\par
\end{samepage}

\paperfigure[!htbp]{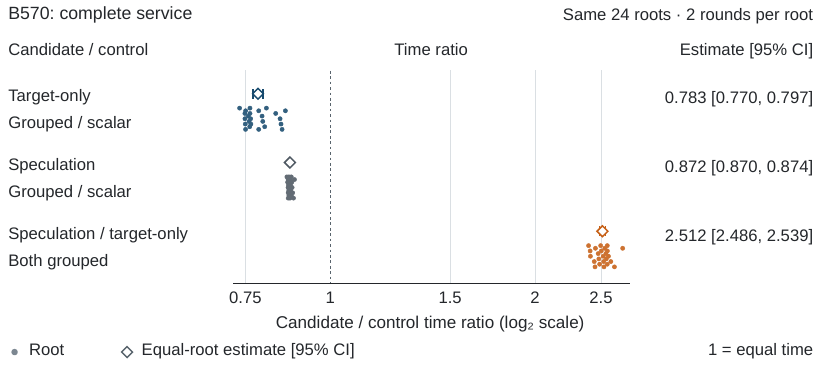}{B570 service with CPU drafting and teardown (24 roots, two rounds). Each contrast reuses the same roots. Open diamonds give equal-root estimates; individual 95\% stratified-root intervals are printed at right. On the common log axis, one means equal time and lower values favor the candidate.}{Each row shows 24 measured root ratios and an open-diamond equal-root summary on one continuous log axis. Grouped/scalar target-only execution is 0.783 with 95\% interval 0.770 to 0.797; grouped/scalar speculation is 0.872 with interval 0.870 to 0.874. Grouped speculation versus grouped target-only is 2.512 with interval 2.486 to 2.539, to the right of the equal-time line at one and therefore slower. The 72 root marks reuse the same 24 roots across three contrasts. The grouping intervals are narrow; their numerical endpoints are printed.}{fig:b570}

For short-lived workers, loading and setup also enter the deployment decision. Tables~\ref{tab:pilot} and~\ref{tab:historical-applications} report the separate draft-placement pilot and worker-lifetime costs.
\FloatBarrier

\section{Related Work}
\label{sec:related}

\paragraph{Stateful contracts and request geometry.}
Vosti's Verus proof links kernel contracts to canonical KV, physical mappings, write isolation and continuation~\cite{vosti}. LOCAL retains admitted projection groups and splits attention by physical view in one decoder call, using inherited quantized kernels.

BlockLock protects a request's geometry as companion requests vary, under repeat-stable arithmetic and exact KV and position handoff~\cite{blocklock}. Both modes share batched prefill: Rows retains batched decode, while Pack extracts the target KV for single-request decode.

\paragraph{Numerical references.}
Yuan et al.\ trace batch- and hardware-dependent output changes to limited-precision arithmetic; their LayerCast pipeline stores weights in BF16 and computes in FP32~\cite{layercast}. LOCAL keeps the inherited quantized profile fixed while admitting projection groups and physical attention views separately. Other interfaces fix a compiled reference and its verified structural class~\cite{taming}. Detent fixes a deployment commitment selected by tuning legal pin values and deployment settings~\cite{detent}; Beyond Accuracy specifies an INT8 operator~\cite{beyondaccuracy}. Batch-invariant kernels remove arithmetic dependencies on other requests~\cite{thinkingmachines}, while TBIK fixes intra- and inter-GPU reduction order across tensor-parallel sizes~\cite{tbik}. FlashInfer's later numerical classes prescribe split-K reductions while permitting tiling and scheduling choices~\cite{flashinferbi}. Realizing these references requires matching hardware rounding~\cite{tensorcoremodels} and accounting for attention layouts and schedules~\cite{flashattention,flashinfer}, as well as implementation-specific reduction or shape choices~\cite{vllm_batchinv,corun}. Tritify distinguishes exact-integer reproducibility from fidelity to a floating-point decision reference~\cite{tritify}.

% Allow this paragraph to continue naturally across the page boundary.
\paragraph{Verification and consumer state.}
Speculative decoding verifies known proposals in parallel~\cite{leviathan2023,chen2023}. State recovery connects verification and commit using verified KV replacement~\cite{llm42,qspec}, full-precision rejection buffers~\cite{quantspec}, or margin-triggered repair~\cite{margingate}. Song~et~al.\ combine fixed-order quantized verification, GPU acceptance and KV compaction for greedy WebGPU decoding~\cite{song2026}. BADC tests processor prefixes, state, stopping behavior and returned records~\cite{badc}. LOCAL executes nonaligned operator partitions, returns owned score rows in logical order, and commits successor state. ESPEON instead verifies completed sequences by deterministic batched recomputation and sampling replay, scoped by layer and token partitions~\cite{espeon}.\par

In the Scientific Supplement, Table~\ref{tab:neighbors} and the source notes in Section~\ref{supp:related-contracts} record the numerical and state interfaces.

\FloatBarrier
\section{Implications for Reference-Preserving Execution}
\label{sec:discussion}

\paragraph{Operator-local constraints.}
The construction allows attention to enforce a physical-view boundary inside a projection group. Applying it to another backend requires specifying that backend's arithmetic recipes, view layout and KV writes, together with the lifetime of the returned scores.

\paragraph{Work recovered at a boundary.}
At the evaluated crossing, SIMPLE already uses few-column arithmetic for q3. LOCAL retains q4 in place of q3+q1 and adds the internal attention views and joins needed to preserve the reference. Other cuts retain different amounts of work. A q32 cut can leave two width-16 groups whose admission and cost depend on the backend profile (Table~\ref{tab:historical-geometry}). In the RTX confirmation, regrouping scalar operations from B to C1 saves verification time but increases mean request time. The descriptive secondary C4 comparisons reduce request time relative to both scalar controls.

\paragraph{Cost over the request lifetime.}
The C4 and P4 gains measure arithmetic and packing choices with SIMPLE fixed. LOCAL removes 20 extra calls across 2,224 logical cycles in M1. In the natural 8B study, shared greedy-sampling differences are comparable to net service differences. With CPU drafting and teardown included, B570 speculative service takes 2.512 times grouped target-only time. For the tested dense-cache, greedy profiles, these observations suggest that a future selector should weigh grouping frequency and avoided work against cost over the request lifetime.

For paged or sliding caches, the next design task is to map each query's reference-visible KV to attention's physical views as pages or windows change. Concurrent execution must bind score rows and cache regions to individual requests and isolate their writes. Stochastic consumers extend the preserved state to include sampler state, binding acceptance, correction and continuation to the same transitions.

\paragraph{Availability.}\label{sec:repro}
The \href{https://github.com/leecdiang/RefBound/tree/de2192a57149150f4fea6811db2c3e8781563e2c}{RefBound companion at commit \texttt{de2192a5}} publishes P4, LOCAL and RTX source, validation records, archived statistical reconstruction and earlier manuscript/figure sources. A5000 LOCAL's new ELF compiled, linked and passed model-free checks; GPU evidence binds the original qualification binary. The rebuilt RTX worker passed device verification without new performance samples. The accompanying delivery provides the qualification manifest separately; Section~\ref{supp:access} maps public files and required original-machine inputs.

\section{Conclusion}
\label{sec:conclusion}

LOCAL preserves projection groups across physical attention-view boundaries. It rejoins attention in query order and returns owned scores. The current call commits the state from which its successor continues. The composition argument gives sufficient conditions for preserving the fixed reference across calls, including operator equivalence on the admitted input domain. In the fixed 32B A5000 study, both arms retain q4 in the primary noncrossing comparison, with an estimated 0.43\% reduction in verification-through-commit time (95\% CI [$-1.13\%$, $2.01\%$]). At the supporting crossing, LOCAL preserves q4 where SIMPLE executes q3+q1, reducing verification-through-commit time by 21.56\% [20.21\%, 22.83\%] over 12 paired time blocks.

\label{page:body-end}
\clearpage
\label{page:references-start}
\bibliographystyle{ACM-Reference-Format}
\bibliography{references}
\label{page:references-end}
\clearpage
\appendix
\setcounter{table}{0}
\setcounter{figure}{0}
\renewcommand{\thefigure}{S\arabic{figure}}
\renewcommand{\theHfigure}{S\arabic{figure}}
\renewcommand{\thetable}{S\arabic{table}}
\renewcommand{\theHtable}{S\arabic{table}}
\section{Scientific Supplement}
\label{sec:supplement}

This supplement gives the statistical roles, direct LOCAL comparisons, separate request studies, natural-service and deployment measurements, and numerical qualifications with their public-access records.

\subsection{Statistical roles and eligibility}
\label{supp:ledger}
Table~\ref{tab:ledger} separates formal observations from warmups, qualification and diagnostics. The estimators in Section~\ref{sec:design} include every eligible measured root; resampling keeps paired rounds within roots and whole time blocks intact. The estimators exclude qualification even when an internal field labels a unit as measured. Resource-marked RTX P4 records and their registered whole-block replacement both remain in the ledger.

A5000 C4 has 12 separate width-qualification requests in addition to its 102-request timing ledger (114 combined). The original LOCAL study retains all eight formal requests; qualification and diagnostic replacements stay outside estimation. B570's 384-request timing table excludes 108 qualification requests and earlier technical attempts; the campaign records 514 launches, 513 completions and one failure. Its pilot adds 15 qualification/capacity requests to eight measured requests and eight warmups, then stops at the predeclared complete-service criterion before the planned 24-root continuation.

\begingroup
\setlength{\intextsep}{3pt plus 1pt minus 1pt}
\begin{table}[H]\caption{Request roles and eligibility for the reported timing estimates.}\label{tab:ledger}% Maintained native LaTeX body | tab:ledger | inherited values; current numbering is automatic.
\begingroup
\fontsize{8.7}{10.5}\selectfont
\setlength{\tabcolsep}{3.5pt}\renewcommand{\arraystretch}{1.03}
\noindent
\begin{tabularx}{\linewidth}{@{}>{\raggedright\arraybackslash}Xrrrr@{}}
\toprule
Study or record set & Measured & Warm / primer & Other & Ledger rows\\
\midrule
\multicolumn{5}{@{}l}{\textbf{RTX and A5000 request ledgers}}\\
RTX P4 & $192$ & $192$ & $8$ & $392$\\
RTX C4 bridge & $144$ & $144$ & $24$ & $312$\\
RTX confirmation & $864$ & $864$ & $36$ & $1{,}764$\\
A5000 confirmation & $96$ & $96$ & $20$ & $212$\\
A5000 P4 & $192$ & $144$ & $48$ & $384$\\
A5000 C4 & $48$ & $48$ & $6$ & $102$\\
\midrule
\multicolumn{5}{@{}l}{\textbf{B570 timing tables}}\\
B570 formal timing table & $192$ & $192$ & $0$ & $384$\\
B570 pilot timing table & $8$ & $8$ & $0$ & $16$\\
\midrule
\multicolumn{5}{@{}l}{\textbf{Reservations and native receipts}}\\
LOCAL reservations & $8$ & $8$ & $11$ & $27$\\
M1 native receipts & $256$ & $32$ & $0$ & $288$\\
\bottomrule
\end{tabularx}
\par\smallskip
\fontsize{8}{9.6}\selectfont\raggedright
\textit{Role definitions.} Other roles: RTX P4 has eight resource-marked rows; initial RTX C4 has 24 qualification rows. RTX confirmation has 24 eligible qualification and 12 resource-excluded rows; its 864 formal measured rows retain the registered complete-six-request replacements. A5000 confirmation has 12 qualification and eight diagnostic units; 48 of its measured units are two-turn episodes. A5000 P4/C4 have 48/6 full-trace captures. LOCAL has six qualification and five diagnostic reservations. No LOCAL formal request was replaced.
\par\endgroup
\end{table}
\endgroup

\subsection{A5000 confirmation: time blocks, endpoints and state}
\label{supp:confirmation}
Confirmation A repeats the selected 32B histories under the original numerical build, with a 1024-slot F16 KV cache (C1024/F16), C4 on, P4 off and frozen proposals. Noncrossing is the primary endpoint; crossing and paired interaction are supporting endpoints (Table~\ref{tab:confirmation-effects}). Confirmation B retains the original 8B Q8\_0 target-only two-turn consumer. Its primary endpoint gives equal log weight to the two fixed conversations. Each confirmation has 12 paired time blocks, analyzed separately from the original study.

\begin{table}[tbp]\caption{All six A5000 confirmation endpoints and their prespecified roles. Direct time ratios are \mbox{LOCAL/SIMPLE}; the supporting interaction divides the crossing geometric-mean ratio by the noncrossing ratio.}\label{tab:confirmation-effects}% Maintained native LaTeX body | tab:confirmation-effects | inherited values; current numbering is automatic.
\begingroup
\fontsize{8.7}{10.5}\selectfont
\setlength{\tabcolsep}{3.5pt}\renewcommand{\arraystretch}{1.03}
\noindent
\begin{tabularx}{\linewidth}{@{}>{\raggedright\arraybackslash}Xlrrr@{}}
\toprule
Endpoint & Role & Ratio & \multicolumn{2}{c}{95\% interval}\\
\cmidrule(lr){4-5}
 &  &  & Lower & Upper\\
\midrule
\multicolumn{5}{@{}l}{\textbf{A / 32B verification through commit}}\\
Noncrossing & Primary & $0.99572472$ & $0.97993067$ & $1.01131242$\\
Crossing & Supporting & $0.78435840$ & $0.77170242$ & $0.79786960$\\
Crossing / noncrossing & Supporting & $0.78772615$ & $0.76779006$ & $0.80563563$\\
\midrule
\multicolumn{5}{@{}l}{\textbf{B / 8B two-turn response plus reset}}\\
Equal-conversation overall & Primary & $0.98555056$ & $0.96147494$ & $1.00824356$\\
Conversation 0 & Supporting & $0.98893282$ & $0.96484815$ & $1.01175466$\\
Conversation 1 & Supporting & $0.98217986$ & $0.94488215$ & $1.01479972$\\
\bottomrule
\end{tabularx}
\par\smallskip
\fontsize{8}{9.6}\selectfont\raggedright
\textit{Note.} All ratios use the confirmation study alone. Direct effects are LOCAL/SIMPLE; A interaction divides the crossing geometric-mean LOCAL/SIMPLE ratio by the noncrossing ratio; below one favors a larger crossing reduction. Each estimate uses 12 joint time blocks. Individual 95\% intervals retain the frozen block-bootstrap draws and endpoint roles, conditional on the observed time window.
\par\endgroup
\end{table}

For confirmation A, let $\ell_{hb}=\log(T^{\mathrm{LOCAL}}_{hb}/T^{\mathrm{SIMPLE}}_{hb})$ for history $h$ in block $b$. The direct estimate is $\exp(12^{-1}\sum_b\ell_{hb})$. The supporting crossing/noncrossing interaction divides these geometric-mean time ratios:
\[
 \widehat r_{\mathrm{int}}=
 \frac{\exp(12^{-1}\sum_b\ell_{\mathrm{cross},b})}
      {\exp(12^{-1}\sum_b\ell_{\mathrm{noncross},b})}.
\]
A value below one means a larger relative LOCAL reduction at the crossing for these two fixed histories. Confirmation B uses $\exp(12^{-1}\sum_b[\ell_{0b}+\ell_{1b}]/2)$ overall, with single-conversation means as supporting endpoints. The 10,000 bootstrap draws (seed 20261117) resample whole blocks jointly; individual 95\% percentile intervals use linear interpolation.

Table~\ref{tab:confirmation-pairs} retains all 12 paired blocks in each confirmation. All crossing pairs favor LOCAL; five noncrossing pairs favor SIMPLE. In B, seven Conversation~0 pairs, six Conversation~1 pairs and five equal-conversation block aggregates favor SIMPLE. Block starts are at least 1,978 seconds apart, with eleven before and one after midnight in \mbox{Asia/Shanghai}. The intervals describe the fixed cases within this observation window. For confirmation A, the geometric-mean LOCAL/SIMPLE time ratios in the first and last six blocks are 0.787 and 0.782 at the crossing, and 0.984 and 1.007 for noncrossing. B gives the conversations equal log weight; the older natural study uses a ratio of total times.

At A's crossing, SIMPLE makes q3+q1 decoder calls and LOCAL makes one q4 call with two attention runs of widths 3 and 1. The successor's 1+3 partition triggers the full-vector guard. The noncrossing successor reuses the GGML graph, while CUDA graphs remain disabled. Mean crossing time falls by 21.555\,ms. The signed differences are $-21.731$\,ms in decode, $+0.176$\,ms in other host work and $+0.000513$\,ms in trim/synchronization. The dependent continuation separately saves 20.858\,ms. Outside this clock, LOCAL incurs 4.924\,ms more prefix-preparation time and 62.252\,ms more study-wrapper time than SIMPLE. The wrapper includes numerical checks at unequal capture depths, so its time does not describe an application endpoint. The original two-repeat study supplies the separate projection/attention device-event diagnostics (Section~\ref{supp:local}).

B retains the parser, wire/client history and re-tokenization chain, a 4096-slot cache (C4096), and two freely generated greedy turns capped at 128 tokens each. The two prompts contain 92 and 250 tokens for Conversation~0, and 27 and 551 for Conversation~1. Their second turns reuse 88 and 23 tokens, respectively, leaving 162 and 528 new prompt tokens. Every turn emits 128 tokens. Conversation~0 has 263 physical calls and 508 target rows in either arm. Conversation~1 processes 809 target rows in 273 SIMPLE calls or 272 LOCAL calls. On this Q8 natural path, LOCAL's 32-query calls beginning at logical positions 247 and 503 cross the 256- and 512-token boundaries, respectively. Each call has two attention runs of widths 9 and 23. Table~\ref{tab:confirmation-costs} gives the response-plus-reset phase differences. CPU reconstruction from the saved native records reproduces all 96 formal A/B values exactly and the paired phase summaries within $1.78\times10^{-15}$\,s. In B, the outer greedy-sampling timer exceeds the sum of its per-token timers by 1.276--1.872\,ms per episode because the outer timer also includes registration of the token-cost record. The service endpoint retains that entire stage.

\begin{table}[!htbp]\caption{Every A5000 confirmation pair. (A) The 32B first-transaction clock measures verification through commit in milliseconds; noncrossing is primary and crossing supporting. (B) The 8B clock measures a complete two-turn response plus reset in seconds.}\label{tab:confirmation-pairs}% Combined confirmation pairs. Values unchanged from s15_confirmation_A and s16_confirmation_B.
\begingroup
\fontsize{8.7}{10.5}\selectfont
\setlength{\tabcolsep}{3.5pt}\renewcommand{\arraystretch}{1.03}
\noindent
\begin{tabular*}{\linewidth}{@{\extracolsep{\fill}}lrrrrrr@{}}
\toprule
\multicolumn{7}{@{}l}{\textbf{A. 32B selected transactions / verification through commit (ms)}}\\
Block & \multicolumn{2}{c}{Noncrossing / base 137} & \multicolumn{2}{c}{Crossing / base 253} & \multicolumn{2}{c}{LOCAL / SIMPLE}\\
\cmidrule(lr){2-3}\cmidrule(lr){4-5}\cmidrule(lr){6-7}
 & SIMPLE & LOCAL & SIMPLE & LOCAL & Noncrossing & Crossing\\
\midrule
$1$ & $73.557$ & $76.023$ & $97.867$ & $75.163$ & $1.033529$ & $0.768008$\\
$2$ & $78.949$ & $74.339$ & $101.913$ & $78.803$ & $0.941608$ & $0.773246$\\
$3$ & $75.613$ & $75.509$ & $99.822$ & $78.979$ & $0.998624$ & $0.791193$\\
$4$ & $76.501$ & $74.641$ & $100.599$ & $80.273$ & $0.975690$ & $0.797947$\\
\addlinespace[2.5pt]
$5$ & $74.716$ & $74.078$ & $99.493$ & $80.558$ & $0.991459$ & $0.809693$\\
$6$ & $77.100$ & $74.594$ & $97.108$ & $75.984$ & $0.967492$ & $0.782470$\\
$7$ & $73.671$ & $73.864$ & $103.343$ & $79.088$ & $1.002621$ & $0.765296$\\
$8$ & $75.867$ & $76.788$ & $99.615$ & $80.113$ & $1.012145$ & $0.804226$\\
\addlinespace[2.5pt]
$9$ & $72.489$ & $73.260$ & $99.116$ & $82.686$ & $1.010628$ & $0.834233$\\
$10$ & $75.461$ & $75.044$ & $100.433$ & $78.269$ & $0.994471$ & $0.779322$\\
$11$ & $74.762$ & $73.169$ & $101.235$ & $77.308$ & $0.978686$ & $0.763648$\\
$12$ & $75.329$ & $78.813$ & $100.171$ & $74.831$ & $1.046254$ & $0.747035$\\
\midrule
\multicolumn{7}{@{}l}{\textbf{B. 8B natural conversations / two-turn response plus reset (s)}}\\
Block & \multicolumn{2}{c}{Conversation 0} & \multicolumn{2}{c}{Conversation 1} & \multicolumn{2}{c}{LOCAL / SIMPLE}\\
\cmidrule(lr){2-3}\cmidrule(lr){4-5}\cmidrule(lr){6-7}
 & SIMPLE & LOCAL & SIMPLE & LOCAL & Conv. 0 & Conv. 1\\
\midrule
$1$ & $9.051$ & $8.940$ & $10.239$ & $10.536$ & $0.987759$ & $1.028995$\\
$2$ & $8.960$ & $9.257$ & $10.180$ & $10.531$ & $1.033146$ & $1.034429$\\
$3$ & $9.323$ & $8.976$ & $10.322$ & $10.686$ & $0.962799$ & $1.035301$\\
$4$ & $9.404$ & $8.560$ & $10.709$ & $10.675$ & $0.910202$ & $0.996834$\\
\addlinespace[2.5pt]
$5$ & $9.354$ & $8.636$ & $11.119$ & $9.777$ & $0.923299$ & $0.879362$\\
$6$ & $9.770$ & $9.816$ & $10.820$ & $9.863$ & $1.004714$ & $0.911589$\\
$7$ & $9.818$ & $9.819$ & $11.095$ & $10.991$ & $1.000046$ & $0.990622$\\
$8$ & $9.826$ & $10.282$ & $10.992$ & $11.139$ & $1.046416$ & $1.013410$\\
\addlinespace[2.5pt]
$9$ & $9.895$ & $10.158$ & $11.119$ & $11.530$ & $1.026540$ & $1.036979$\\
$10$ & $10.651$ & $10.206$ & $11.525$ & $11.503$ & $0.958194$ & $0.998081$\\
$11$ & $10.142$ & $10.268$ & $11.582$ & $11.861$ & $1.012493$ & $1.024109$\\
$12$ & $10.545$ & $10.672$ & $11.951$ & $10.279$ & $1.012012$ & $0.860096$\\
\bottomrule
\end{tabular*}
\par\smallskip
\fontsize{8}{9.6}\selectfont\raggedright
\textit{Note.} Each panel contains 48 formal arm observations over 12 paired blocks, analyzed separately. Panel A uses milliseconds and Panel B seconds. Ratios use unrounded observations; values above one indicate longer LOCAL time. Conversation numbers identify the two natural workloads.
\par\endgroup
\end{table}

\begin{table}[tbp]\caption{Arithmetic mean LOCAL-minus-SIMPLE phase differences in milliseconds within the registered response-plus-reset clock.}\label{tab:confirmation-costs}% Maintained native LaTeX body | tab:confirmation-costs | inherited values; current numbering is automatic.
\begingroup
\fontsize{8.7}{10.5}\selectfont
\setlength{\tabcolsep}{3.5pt}\renewcommand{\arraystretch}{1.03}
\noindent
\begin{tabularx}{\linewidth}{@{}>{\raggedright\arraybackslash}Xrr@{}}
\toprule
Phase & \multicolumn{2}{c}{LOCAL minus SIMPLE (ms)}\\
\cmidrule(lr){2-3}
 & Conversation 0 & Conversation 1\\
\midrule
\multicolumn{3}{@{}l}{\textbf{Greedy token selection and controller overhead}}\\
Shared greedy-sampling stage & $-112.573$ & $-192.930$\\
Controller reset overhead & $-30.006$ & $+51.671$\\
Controller response overhead & $+68.878$ & $+41.394$\\
\addlinespace[2.5pt]
\multicolumn{3}{@{}l}{\textbf{Execution, state and request accounting}}\\
Consumer transition & $-0.261$ & $-0.545$\\
Canonical history producer & $-0.676$ & $-0.540$\\
Decode, sync and persistent ledger & $-21.794$ & $-79.665$\\
Native reset & $+0.926$ & $-0.877$\\
Native response residual & $-0.437$ & $-0.546$\\
Owned output copy & $+0.572$ & $-6.951$\\
Reconcile and checkpoint & $-0.069$ & $-0.112$\\
Request finish & $-0.243$ & $-0.520$\\
Request reserve & $-0.086$ & $-0.416$\\
\midrule
\textbf{Complete response plus reset} & $-95.769$ & $-190.038$\\
\bottomrule
\end{tabularx}
\par\smallskip
\fontsize{8}{9.6}\selectfont\raggedright
\textit{Clock.} Within the separate 8B natural-service confirmation (Panel B). Mean component differences sum to mean service difference up to rounding; negative values save time. Decode includes synchronization and a persistent per-call ledger. Scientific payload export and independent checking occur outside the original service clock. Arithmetic phase differences and the geometric primary ratio summarize the same service endpoint with different estimators.
\par\endgroup
\end{table}

\par\penalty0\relax
\begin{samepage}
The 212 requests comprise 12 qualification, 96 fresh same-arm warmups, 96 formal observations and eight diagnostics. A and B each contain 48 formal arm observations and 106 total units; a B unit is a two-turn episode. Technical replacements were disabled, and there were no failures, interruptions or reruns. Recorded work is 91,319 target rows and zero draft rows.\par
\end{samepage}

\begin{samepage}
Original-machine captures check 954 A owned/frontier vectors and 230 canonical checkpoints, plus 27,136 B vocabulary vectors and 444 active-KV snapshots. Capture depth varies by role; the totals aggregate the recorded captures across roles. Public numerical access is specified in Table~\ref{tab:access}.\par
\end{samepage}

\subsection{Original selected transactions and dependent successors}
\label{supp:local}
The original LOCAL study uses CodeContests/train/161 137- and 253-token histories, with C4 on, P4 off, frozen proposals and two paired repeats each. Successors begin at 139 and 255 after their own preceding commits. Its verification-through-commit clock includes owned-row copying, decisions, trim and synchronization, and excludes history reconstruction, loading, live drafting and payload persistence. Table~\ref{tab:local-pairs} retains all first-transaction and dependent-successor pairs.

\begin{table}[tbp]\caption{Every original LOCAL pair, including successors from each arm's own committed state.}\label{tab:local-pairs}% Maintained native LaTeX body | tab:local-pairs | inherited values; current numbering is automatic.
\begingroup
\fontsize{8.7}{10.5}\selectfont
\setlength{\tabcolsep}{3.0pt}\renewcommand{\arraystretch}{1.03}
\noindent
\begin{tabular*}{\linewidth}{@{\extracolsep{\fill}}lrrrrrr@{}}
\toprule
Repeat & \multicolumn{3}{c}{Noncrossing} & \multicolumn{3}{c}{Crossing}\\
\cmidrule(lr){2-4}\cmidrule(lr){5-7}
 & SIMPLE (ms) & LOCAL (ms) & L/S & SIMPLE (ms) & LOCAL (ms) & L/S\\
\midrule
\multicolumn{7}{@{}l}{\textbf{First transaction / history bases 137 (noncrossing), 253 (crossing)}}\\
$1$ & $74.147$ & $75.077$ & $1.012542$ & $100.360$ & $77.312$ & $0.770351$\\
$2$ & $79.035$ & $74.999$ & $0.948940$ & $99.638$ & $77.475$ & $0.777572$\\
\midrule
\multicolumn{7}{@{}l}{\textbf{Dependent successor / history bases 139 (noncrossing), 255 (crossing)}}\\
$1$ & $74.512$ & $75.885$ & $1.018418$ & $102.813$ & $83.540$ & $0.812547$\\
$2$ & $77.112$ & $75.688$ & $0.981536$ & $104.247$ & $81.747$ & $0.784163$\\
\bottomrule
\end{tabular*}
\par\smallskip
\fontsize{8}{9.6}\selectfont\raggedright
\textit{Note.} Times are clean verify-through-commit milliseconds; ratio is LOCAL/SIMPLE. Each successor depends on its own first commit. One root, two histories and two paired repeats supply four first-transaction pairs; successors are not additional independent repeats. Unrounded observations remain in the figure data.
\par\endgroup
\end{table}

The crossing first-transaction geometric ratio is 0.77395294, with reductions of 22.96\% and 22.24\% in the two paired repeats. Its successor ratio is 0.79822856. The noncrossing ratios are 0.98022489 for the first transaction (a 1.25\% increase and a 5.11\% reduction) and 0.99980702 for the successor. These descriptive fixed-history estimates have no population interval; successors remain dependent observations within their histories.

\begin{table}[tbp]\caption{Separate device-event diagnostics from the original two-repeat transaction study. Each profiler run measures one first-transaction history; times and counts use its own clock.}\label{tab:diagnostics}% Maintained native LaTeX body | tab:diagnostics | inherited values; current numbering is automatic.
\begingroup
\fontsize{8.7}{10.5}\selectfont
\setlength{\tabcolsep}{3.5pt}\renewcommand{\arraystretch}{1.03}
\noindent
\begin{tabularx}{\linewidth}{@{}>{\raggedright\arraybackslash}Xrrrr@{}}
\toprule
Profiler output & \multicolumn{2}{c}{Noncrossing / base 137} & \multicolumn{2}{c}{Crossing / base 253}\\
\cmidrule(lr){2-3}\cmidrule(lr){4-5}
 & SIMPLE & LOCAL & SIMPLE & LOCAL\\
\midrule
\multicolumn{5}{@{}l}{\textbf{Event counts}}\\
Projection prepare calls & $449$ & $449$ & $834$ & $449$\\
Projection dot calls & $449$ & $449$ & $834$ & $449$\\
Attention events & $192$ & $192$ & $384$ & $384$\\
Join calls & $0$ & $0$ & $0$ & $64$\\
\midrule
\multicolumn{5}{@{}l}{\textbf{Device-event duration (ms)}}\\
Projection & $53.695$ & $53.794$ & $74.334$ & $53.417$\\
Attention & $9.127$ & $9.090$ & $12.060$ & $11.621$\\
Joins & $0$ & $0$ & $0$ & $0.134$\\
Memcpy aggregate & $0.268$ & $0.266$ & $0.361$ & $0.494$\\
\bottomrule
\end{tabularx}
\par\smallskip
\fontsize{8}{9.6}\selectfont\raggedright
\textit{Clock.} Each column is a separate profiler run from the original two-repeat transaction study. The crossing SIMPLE preparation and dot categories each contain 449 q3 plus 385 q1 events; LOCAL has 449 q4. The q1 path partly uses gate/up fusion. Event counts do not count fused operators. Projection time sums prepare and dot; memcpy is the exported aggregate. The device-event sums and clean wall clock are reported separately.
\par\endgroup
\end{table}

Table~\ref{tab:diagnostics} separates projection, attention, joins and exported memcpy costs. The diagnostic timings may include profiler and cold-call overhead. They cover the full lowering described in Section~\ref{sec:method}.

% Keep original-transaction tables before the separate RTX confirmation.
\FloatBarrier
\subsection{RTX confirmation after shared activation preparation}
\label{supp:rtx-confirmation}

The RTX confirmation reuses the numerical core, native workers, models and 24 previously used roots from the initial B/C1/C4 study. Six time blocks each contain two rounds per root and three arms, giving 12 rounds per root. Every root receives each of the six arm orders twice, with eight observations per domain/block/arm/position cell. Each root-round uses three fresh workers with same-arm warmups. The campaign runs from 17 November 2026 21:33 to 18 November 06:56 in \mbox{Asia/Shanghai}, with block starts at least 1,805.7 seconds apart.

C1/B is the sole prespecified primary comparison; C4/C1 and C4/B are descriptive secondary contrasts (Table~\ref{tab:rtx269-effects}). For root $i$, block $b$ and its two rounds $j$,
\begin{equation}
 \widehat r_{\mathrm{RTX}}=\exp\!\left[\frac{1}{24\cdot6}\sum_{i=1}^{24}\sum_{b=1}^{6}\left(\frac12\sum_{j=1}^{2}\log\frac{T^{\mathrm{candidate}}_{ibj}}{T^{\mathrm{control}}_{ibj}}\right)\right].
 \label{eq:rtx-estimator}
\end{equation}
The 10,000 prespecified draws independently resample roots within balanced code/math strata and whole time blocks; all contrasts share draws and use linear percentile endpoints. C1/B increases time by an estimated 0.90\%, with an interval from a 0.48\% reduction to a 2.19\% increase. The upper endpoint of 2.19\% exceeds the prespecified $\pm2\%$ practical interpretation band.

\begin{table}[tbp]\caption{RTX confirmation primary and descriptive complete-request estimates, with crossed root/time-block intervals.}\label{tab:rtx269-effects}% Maintained native LaTeX body | tab:rtx269-effects | inherited values; current numbering is automatic.
\begingroup
\fontsize{8.7}{10.5}\selectfont
\setlength{\tabcolsep}{3.0pt}\renewcommand{\arraystretch}{1.03}
\noindent
\begin{tabularx}{\linewidth}{@{}>{\raggedright\arraybackslash}Xlrrrr@{}}
\toprule
Candidate / control & Registered role & Time ratio & \multicolumn{2}{c}{95\% ratio interval} & Reduction (\%)\\
\cmidrule(lr){4-5}
 &  &  & Lower & Upper & \\
\midrule
C1/B & Primary & $1.009014$ & $0.995205$ & $1.021886$ & $-0.90$\\
\midrule
C4/C1 & Descriptive & $0.842031$ & $0.816969$ & $0.869273$ & $+15.80$\\
C4/B & Descriptive & $0.849622$ & $0.825148$ & $0.876047$ & $+15.04$\\
\bottomrule
\end{tabularx}
\par\smallskip
\fontsize{8}{9.6}\selectfont\raggedright
\textit{Note.} Roots receive equal weight, as do the six time blocks; two paired log ratios are averaged within each root/block. Intervals use the 10,000 prespecified crossed root/time-block draws. Negative reduction denotes longer time.
\par\endgroup
\end{table}

The conditional root-only C1/B interval is [1.001432, 1.015407], holding the six blocks fixed. Tables~\ref{tab:rtx269-blocks} and~\ref{tab:rtx269-roots} report every block and root. One C1/B block favors C1 and five favor B; all 24 root aggregates and all six block aggregates favor C4 against either scalar control.

\begin{table}[tbp]\caption{All six RTX time blocks in execution order: within-block geometric time ratios.}\label{tab:rtx269-blocks}% Maintained native LaTeX body | tab:rtx269-blocks | inherited values; current numbering is automatic.
\begingroup
\fontsize{8.7}{10.5}\selectfont
\setlength{\tabcolsep}{3.0pt}\renewcommand{\arraystretch}{1.03}
\noindent
\begin{tabularx}{\linewidth}{@{}>{\raggedright\arraybackslash}Xrrrrrr@{}}
\toprule
Comparison & \multicolumn{6}{c}{Time block in execution order}\\
\cmidrule(lr){2-7}
Time ratio & $1$ & $2$ & $3$ & $4$ & $5$ & $6$\\
\midrule
C1/B & $1.008157$ & $0.982917$ & $1.022680$ & $1.009428$ & $1.016651$ & $1.014735$\\
C4/C1 & $0.853166$ & $0.848897$ & $0.845856$ & $0.836798$ & $0.836733$ & $0.830954$\\
C4/B & $0.860125$ & $0.834395$ & $0.865041$ & $0.844687$ & $0.850666$ & $0.843198$\\
\bottomrule
\end{tabularx}
\par\smallskip
\fontsize{8}{9.6}\selectfont\raggedright
\textit{Note.} Each cell gives equal log weight to 24 roots and their two paired rounds within that block. These six temporal summaries retain all 864 formal measured requests from the same input panel.
\par\endgroup
\end{table}
\begin{table}[tbp]\caption{All 24 RTX roots: per-root geometric time ratios over 12 rounds.}\label{tab:rtx269-roots}% Maintained native LaTeX body | tab:rtx269-roots | inherited values; current numbering is automatic.
\begingroup
\fontsize{8.5}{10.2}\selectfont
\setlength{\tabcolsep}{2.5pt}\renewcommand{\arraystretch}{1.03}
\noindent
\begin{tabularx}{\linewidth}{@{}lrrr@{\hspace{9pt}}>{\raggedright\arraybackslash}Xrrr@{}}
\toprule
\multicolumn{4}{c}{CodeContests / test} & \multicolumn{4}{c}{MATH / train}\\
\cmidrule(lr){1-4}\cmidrule(lr){5-8}
ID & C1/B & C4/C1 & C4/B & Category / ID & C1/B & C4/C1 & C4/B\\
\midrule
$116$ & $0.977393$ & $0.893160$ & $0.872969$ & counting/prob. / 35 & $1.019006$ & $0.851442$ & $0.867625$\\
$81$ & $1.002746$ & $0.888316$ & $0.890755$ & geometry / 701 & $1.005114$ & $0.886351$ & $0.890884$\\
$124$ & $0.943621$ & $0.989180$ & $0.933411$ & prealgebra / 316 & $1.013964$ & $0.909168$ & $0.921863$\\
$60$ & $1.013571$ & $0.883379$ & $0.895367$ & geometry / 545 & $1.029702$ & $0.854413$ & $0.879791$\\
\addlinespace[2.5pt]
$108$ & $0.993565$ & $0.885871$ & $0.880170$ & algebra / 355 & $1.020987$ & $0.862575$ & $0.880678$\\
$129$ & $0.993481$ & $0.885671$ & $0.879897$ & algebra / 1050 & $1.021277$ & $0.903237$ & $0.922455$\\
$52$ & $1.013151$ & $0.857185$ & $0.868457$ & counting/prob. / 398 & $1.017737$ & $0.731980$ & $0.744963$\\
$159$ & $1.015929$ & $0.825966$ & $0.839123$ & algebra / 320 & $1.013018$ & $0.731972$ & $0.741500$\\
\addlinespace[2.5pt]
$0$ & $1.015562$ & $0.832252$ & $0.845204$ & geometry / 387 & $1.015056$ & $0.837771$ & $0.850385$\\
$71$ & $1.022598$ & $0.822217$ & $0.840798$ & algebra / 337 & $1.015701$ & $0.721457$ & $0.732785$\\
$90$ & $1.028435$ & $0.911259$ & $0.937171$ & algebra / 917 & $0.997406$ & $0.720716$ & $0.718847$\\
$6$ & $1.023586$ & $0.836972$ & $0.856714$ & geometry / 567 & $1.007740$ & $0.753217$ & $0.759047$\\
\bottomrule
\end{tabularx}
\par\smallskip
\fontsize{8}{9.6}\selectfont\raggedright
\textit{Note.} The two lists contain distinct code and math roots under common column headings. Root ratios retain all 12 rounds with equal block weighting. Code denotes CodeContests; counting/prob. denotes the MATH counting\_and\_probability category. The earlier bridge used the same input panel and is analyzed separately.
\par\endgroup
\end{table}

\begin{samepage}
Prefill and drafting offset C1's verification saving relative to B: the phase differences sum to $+80.52$\,ms per request. C4/C1 saves 1,411.60\,ms overall, including a 1,465.29\,ms verification saving and 63.21\,ms added prefill time (Table~\ref{tab:rtx269-phases}). An offline audit of all 864 formal records reproduces these costs and the root/block directions. Paired work counts match. The ten nonoverlapping stages sum to the service time within $3.56\times10^{-15}$\,s.\par
\end{samepage}

\begin{table}[tbp]\caption{RTX confirmation arithmetic phase differences, candidate minus control, within the complete resident-request clock.}\label{tab:rtx269-phases}% Maintained native LaTeX body | tab:rtx269-phases | inherited values; current numbering is automatic.
\begingroup
\fontsize{8.7}{10.5}\selectfont
\setlength{\tabcolsep}{3.5pt}\renewcommand{\arraystretch}{1.03}
\noindent
\begin{tabularx}{\linewidth}{@{}>{\raggedright\arraybackslash}Xrrr@{}}
\toprule
Phase & \multicolumn{3}{c}{Candidate minus control (ms)}\\
\cmidrule(lr){2-4}
 & C1 $-$ B & C4 $-$ C1 & C4 $-$ B\\
\midrule
\multicolumn{4}{@{}l}{\textbf{Target work}}\\
Target prefill & $+173.574$ & $+63.214$ & $+236.788$\\
Target decode & $+0.000$ & $+0.000$ & $+0.000$\\
Target verification & $-100.426$ & $-1465.286$ & $-1565.712$\\
\addlinespace[2.5pt]
\multicolumn{4}{@{}l}{\textbf{Draft work}}\\
Draft prefill & $+3.300$ & $+0.409$ & $+3.709$\\
Draft synchronization & $+1.211$ & $-2.540$ & $-1.330$\\
Draft proposals & $+2.921$ & $-7.631$ & $-4.710$\\
\addlinespace[2.5pt]
\multicolumn{4}{@{}l}{\textbf{Request and host work}}\\
Reset/context & $+0.016$ & $-0.024$ & $-0.008$\\
Decision/commit & $-0.082$ & $+0.388$ & $+0.306$\\
Terminal synchronization & $+0.000$ & $-0.000$ & $+0.000$\\
Other host & $+0.002$ & $-0.129$ & $-0.128$\\
\midrule
\textbf{Complete service} & $+80.515$ & $-1411.599$ & $-1331.084$\\
\midrule
Verification faster: roots / blocks & 24/24; 6/6 & 24/24; 6/6 & 24/24; 6/6\\
Service faster: roots / blocks & 5/24; 1/6 & 24/24; 6/6 & 24/24; 6/6\\
\bottomrule
\end{tabularx}
\par\smallskip
\fontsize{8}{9.6}\selectfont\raggedright
\textit{Clock.} The C1 minus B column has the opposite sign to Figure~\ref{fig:c4}c, which shows B minus C1 time differences. A negative service value in the figure means that C1 takes longer. Positive values add time to the candidate. The nonoverlapping phase differences sum to the native-service time difference before rounding. Geometric ratios in Table~\ref{tab:rtx269-effects} have a different estimand. Direction counts use the mean paired time difference within each root (12 pairs) or block (48 pairs); roots and blocks describe the same observations. All 72 root and 18 block cost rows accompany the revision.
\par\endgroup
\end{table}

C4 records 2,273,400 verification admissions and 2,652 prefill admissions. All prefill admissions occur at the q3 tail of the 163-token MATH/counting\_and\_probability/train/398 prompt, in its 12 measured requests. B and C1 have zero few-column admissions. The treatment includes both target verification and that observed prefill exposure. Prompts contain 140--757 tokens and visible outputs contain 42--128. Including end-of-generation (EOG) tokens, decision sequences have 43--128 tokens. The ledger contains 684 cap-terminated requests and 180 that end with EOG. Separate initial-EOG and complete logical q1-tail qualifications remain outside this study.

The 1,764-row ledger separates 864 formal observations, 864 warmups, 24 eligible qualification and 12 resource-excluded requests (Table~\ref{tab:ledger}). The excluded units are one six-request qualification unit and one six-request formal root-round, replaced under the registered allowance with their original records retained. Qualification contains twelve capture warmups and twelve clean requests; neither enters estimation.

Qualification on the original machine compares fresh candidate captures with stored reference captures on two sentinel roots. The overlapping counts are 1,932 comparisons with stored target-reference vectors, 3,768 target/draft vector comparisons and 456 canonical checkpoints. Every timed request includes checks of complete-state/frontier identities, pending and transaction state, actual target/draft work, dispatch and graph/view identity.

\begin{samepage}
The later fixed-input C1/C4 capture independently identifies the verification kernels and checks complete intermediate KV and continuation (Table~\ref{tab:access}). The released scalar timings and bootstrap draws reconstruct these estimates, and the role register identifies the primary comparison and descriptive contrasts.
\par
\end{samepage}

\subsection{Complete-request effects and phase accounts}
\label{supp:effects}
The initial request studies use Eq.~\eqref{eq:estimator}: paired log ratios are averaged over a root's rounds before giving roots equal weight. Whole-root bootstrap retains rounds within code/math strata. The 10,000-draw percentile intervals use linear interpolation, seed 20261117 for A5000 C4 and 20261116 for the other modern panels. Table~\ref{tab:direct-effects} includes every direct contrast from the five principal application studies.

\par\penalty0\relax
\begin{samepage}
A5000 has 232 P4-entering prefill calls across TO\_P4 and SD\_P4, all aligned q32 within one physical extent. Each RTX P4 arm records 140,114 q32, 442 q8 and 2,652 other-width projection dispatches, with zero q16 in the recorded buckets: 143,208 dispatches per arm. Dispatch counts and outer-call counts are distinct units.\par
\end{samepage}

\begin{table}[tbp]\caption{All direct modern request effects, retaining each separate study and its contemporaneous control.}\label{tab:direct-effects}% Maintained native LaTeX body | tab:direct-effects | inherited values; current numbering is automatic.
\begingroup
\fontsize{8.7}{10.5}\selectfont
\setlength{\tabcolsep}{3.2pt}\renewcommand{\arraystretch}{1.03}
\noindent
\begin{tabularx}{\linewidth}{@{}>{\raggedright\arraybackslash}Xllrrr@{}}
\toprule
Study & Candidate & Control & Time ratio & \multicolumn{2}{c}{95\% interval}\\
\cmidrule(lr){5-6}
 &  &  &  & Lower & Upper\\
\midrule
A5000 C4 & C4 & C1 & $0.722314$ & $0.699108$ & $0.745748$\\
\midrule
RTX C4 (initial) & C4 & C1 & $0.844588$ & $0.815651$ & $0.873844$\\
 & C4 & B & $0.851523$ & $0.822198$ & $0.881192$\\
 & C1 & B & $1.008211$ & $0.993614$ & $1.022398$\\
\midrule
A5000 P4 & TO\_P4 & TO\_C4 & $0.868192$ & $0.842744$ & $0.896176$\\
 & SD\_P4 & SD\_C4 & $0.850524$ & $0.822399$ & $0.882039$\\
 & SD\_P4 & TO\_P4 & $0.864045$ & $0.841128$ & $0.887396$\\
 & SD\_P4 & TO\_C4 & $0.750157$ & $0.718747$ & $0.784926$\\
\midrule
RTX P4 & TO\_P4 & TO\_C4 & $0.538107$ & $0.506846$ & $0.572833$\\
 & SD\_P4 & SD\_C4 & $0.504723$ & $0.481321$ & $0.529974$\\
 & SD\_P4 & TO\_P4 & $0.850891$ & $0.825631$ & $0.875029$\\
 & SD\_P4 & TO\_C4 & $0.457870$ & $0.441808$ & $0.474843$\\
\midrule
B570 formal & TO\_C & TO\_A & $0.782728$ & $0.769868$ & $0.796538$\\
 & SD\_C & SD\_A & $0.872182$ & $0.870051$ & $0.874405$\\
 & SD\_C & TO\_C & $2.512087$ & $2.485677$ & $2.538986$\\
 & SD\_C & TO\_A & $1.966281$ & $1.942815$ & $1.990796$\\
\bottomrule
\end{tabularx}
\par\smallskip
\fontsize{8}{9.6}\selectfont\raggedright
\textit{Note.} Equal-root geometric paired effects from Eq.~\ref{eq:estimator}; individual 95\% stratified-root bootstrap intervals. Every contrast retains its own contemporaneous denominator. The editable data contain every root, arm and round.
\par\endgroup
\end{table}

\paperfigure{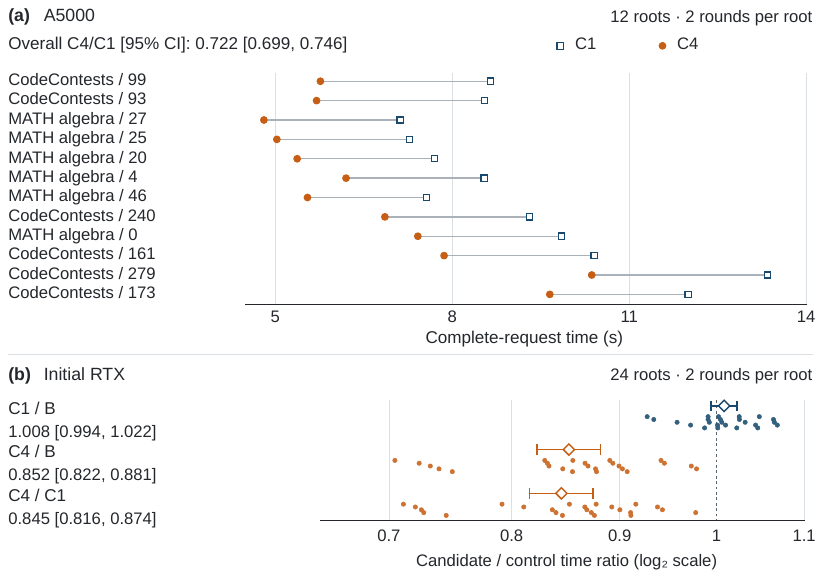}{Initial C4 studies with SIMPLE fixed and P4 off, separate from the 12-round RTX confirmation. (a) Twelve A5000 roots: paired C1/C4 geometric complete-request seconds over two rounds. (b) Twenty-four initial RTX roots: geometric time ratios for C1/B, C4/B and C4/C1 over two rounds, with equal time at one. Summaries give equal-root estimates and individual 95\% stratified-root intervals. Each platform uses its own paired controls.}{The upper panel uses seconds and links the C1 open square to the C4 filled circle for each of 12 A5000 roots. The lower panel uses time ratios for 24 initial RTX roots in each of three direct comparisons. C1/B has an estimate of 1.008 and interval 0.994 to 1.022, crossing equal time; both C4 comparisons lie below one. Each initial study has two rounds, its own paired control and its own interval; the later RTX confirmation has a separate six-block estimate.}{fig:c4-initial}

\par\penalty0\relax
\begin{samepage}
Table~\ref{tab:phase-costs} reports nonoverlapping arithmetic mean phase differences, including residual host cost, that sum to the mean time difference at each named service boundary. These differences and geometric effects have separate estimands. CUDA resident clocks exclude model/context setup and post-clock exports; B570 includes normal teardown.\par
\end{samepage}

\begin{table}[tbp]\caption{Signed arithmetic mean phase differences in milliseconds. Each column is one same-study SD/SD comparison.}\label{tab:phase-costs}% Maintained native LaTeX body | tab:phase-costs | inherited values; current numbering is automatic.
\begingroup
\fontsize{8.7}{10.5}\selectfont
\setlength{\tabcolsep}{3.5pt}\renewcommand{\arraystretch}{1.03}
\noindent
\begin{tabularx}{\linewidth}{@{}>{\raggedright\arraybackslash}Xrrrr@{}}
\toprule
Request component & \multicolumn{2}{c}{A5000 studies} & \multicolumn{2}{c}{RTX studies}\\
\cmidrule(lr){2-3}\cmidrule(lr){4-5}
Difference (ms) & C4 $-$ C1 & P4 $-$ C4 & C4 $-$ C1 & P4 $-$ C4\\
\midrule
Target prefill & $+3.787$ & $-1064.536$ & $+71.206$ & $-4456.997$\\
Target verification & $-2462.096$ & $-5.679$ & $-1457.116$ & $-11.742$\\
Draft: prefill / sync / proposal & $-28.682$ & $-9.986$ & $-10.401$ & $-3.555$\\
Decision and commit & $+9.794$ & $+3.143$ & $+0.342$ & $+0.116$\\
Context / final sync / host residual & $-0.029$ & $-0.026$ & $-0.155$ & $-0.012$\\
\midrule
Complete resident request & $-2477.226$ & $-1077.084$ & $-1396.124$ & $-4472.190$\\
\bottomrule
\end{tabularx}
\par\smallskip
\fontsize{8}{9.6}\selectfont\raggedright
\textit{Note.} C4 $-$ C1 compares the speculative C4 and C1 arms; P4 $-$ C4 compares SD\_P4 and SD\_C4. Arithmetic phase means within each named study; negative values save time. Target decode is zero in these SD/SD comparisons. The nonoverlapping phase differences sum to each study's resident-request time difference, up to rounding. TTFT is not an additive phase. These costs are separate from the geometric paired effects and are not cumulative stages across studies.
\par\endgroup
\end{table}

% Keep the preceding request-effect figure and tables before M1.
\FloatBarrier
\subsection{Natural service and the complete M1 design}
\label{supp:natural}
M1 uses the historical name WHOLE for the whole-call splitting baseline. It crosses WHOLE/LOCAL with loop4/packed projection organization, retaining four sessions within each of 16 roots. Every arm has 2,224 logical cycles. WHOLE executes 2,244 physical verification calls: 2,204 q4, 16 q2, 12 q1 and 12 q3. LOCAL executes 2,224 q4 calls, removing 20 extra calls (0.8993\% of cycles).

\begin{table}[!htbp]\caption{M1's five-comparison family. Direct contrasts use ratios of totals; interaction uses root-geometric ratios.}\label{tab:m1-family}% Maintained native LaTeX body | tab:m1-family | inherited values; current numbering is automatic.
\begingroup
\fontsize{8.7}{10.5}\selectfont
\setlength{\tabcolsep}{3.5pt}\renewcommand{\arraystretch}{1.03}
\noindent
\begin{tabularx}{\linewidth}{@{}>{\raggedright\arraybackslash}Xlrrr@{}}
\toprule
Comparison & Held fixed & Ratio & \multicolumn{2}{c}{99\% interval}\\
\cmidrule(lr){4-5}
 &  &  & Lower & Upper\\
\midrule
LOCAL/WHOLE & loop4 & $0.99791268$ & $0.99536975$ & $0.99957575$\\
LOCAL/WHOLE & packed & $0.99871654$ & $0.99595261$ & $1.00056974$\\
Packed/loop4 & WHOLE & $0.98364752$ & $0.98023586$ & $0.98865441$\\
Packed/loop4 & LOCAL & $0.98443989$ & $0.98122828$ & $0.98939096$\\
\midrule
Root-geometric interaction & --- & $1.00069601$ & $1.00006926$ & $1.00139281$\\
\bottomrule
\end{tabularx}
\par\smallskip
\fontsize{8}{9.6}\selectfont\raggedright
\textit{Family.} The first four estimates are ratios of native-service totals, retaining four sessions per root. Interaction is the root-geometric mean of $(L_P/W_P)/(L_{L4}/W_{L4})$ and exceeds one when packing attenuates the LOCAL effect. Each interval is a 99\% marginal member of the prespecified five-comparison Bonferroni family, conditional on the four observed sessions.
\par\endgroup
\end{table}

Let $T_{a,i}$ sum arm $a$'s time over root $i$'s four sessions. Each direct effect is the ratio of totals, $\sum_i T_{\mathrm{candidate},i}/\sum_i T_{\mathrm{control},i}$; paired-root resampling retains all four sessions. The interaction instead averages root-level log ratios, where $L$ and $W$ denote LOCAL and WHOLE, and subscripts $P$ and $L4$ denote packed and loop4:
\[
 \exp\!\left[\frac{1}{16}\sum_i\log
 \frac{L_{P,i}/W_{P,i}}{L_{L4,i}/W_{L4,i}}\right].
\]
The historical bootstrap uses 10,000 paired-root draws (seed 19720261107) and 99\% marginal intervals for the five-comparison Bonferroni family. RefBound recomputes the four arm totals, direct point estimates and root-geometric interaction from 256 measured requests, retaining 32 primers as excluded records. The public script retains the original 99\% interval endpoints as frozen historical metadata (\texttt{intervals\_recomputed=false}). Relative to WHOLE, LOCAL takes longer on five loop4 roots and ten packed roots. Packing reduces time on every root compared with loop4 under either cut policy (Table~\ref{tab:m1-roots}). The interaction ratio of 1.00070 indicates a small attenuation of LOCAL's effect with packing.

\begin{table}[tbp]\caption{Every M1 paired root. Times sum four sessions in seconds; each row gives its direct cut-policy ratios.}\label{tab:m1-roots}% Maintained native LaTeX body | tab:m1-roots | inherited values; current numbering is automatic.
\begingroup
\fontsize{8.7}{10.5}\selectfont
\setlength{\tabcolsep}{2.8pt}\renewcommand{\arraystretch}{1.03}
\noindent
\begin{tabularx}{\linewidth}{@{}>{\raggedright\arraybackslash}Xrrrrrr@{}}
\toprule
Training root & \multicolumn{3}{c}{loop4} & \multicolumn{3}{c}{packed}\\
\cmidrule(lr){2-4}\cmidrule(lr){5-7}
Category / ID & WHOLE (s) & LOCAL (s) & L/W & WHOLE (s) & LOCAL (s) & L/W\\
\midrule
\multicolumn{7}{@{}l}{\textbf{APPS / train}}\\
$413$ & $6.826$ & $6.724$ & $0.985078$ & $6.739$ & $6.642$ & $0.985482$\\
$475$ & $12.443$ & $12.400$ & $0.996566$ & $12.186$ & $12.141$ & $0.996290$\\
$726$ & $15.234$ & $15.225$ & $0.999432$ & $14.903$ & $14.918$ & $1.000975$\\
$733$ & $10.547$ & $10.544$ & $0.999784$ & $10.339$ & $10.356$ & $1.001615$\\
$845$ & $7.388$ & $7.391$ & $1.000398$ & $7.217$ & $7.218$ & $1.000102$\\
$859$ & $13.166$ & $13.108$ & $0.995643$ & $12.903$ & $12.859$ & $0.996577$\\
$879$ & $9.575$ & $9.577$ & $1.000254$ & $9.401$ & $9.404$ & $1.000247$\\
$914$ & $18.716$ & $18.701$ & $0.999214$ & $18.286$ & $18.306$ & $1.001068$\\
\midrule
\multicolumn{7}{@{}l}{\textbf{MATH / train}}\\
algebra / 313 & $5.470$ & $5.469$ & $0.999885$ & $5.413$ & $5.417$ & $1.000699$\\
geometry / 333 & $5.788$ & $5.785$ & $0.999360$ & $5.741$ & $5.737$ & $0.999295$\\
geometry / 654 & $5.897$ & $5.899$ & $1.000252$ & $5.841$ & $5.845$ & $1.000604$\\
interm. algebra / 315 & $6.902$ & $6.848$ & $0.992225$ & $6.820$ & $6.766$ & $0.992042$\\
number theory / 523 & $7.152$ & $7.152$ & $1.000003$ & $7.088$ & $7.082$ & $0.999156$\\
prealgebra / 1186 & $5.612$ & $5.612$ & $1.000048$ & $5.555$ & $5.556$ & $1.000199$\\
prealgebra / 142 & $5.599$ & $5.586$ & $0.997664$ & $5.578$ & $5.581$ & $1.000570$\\
prealgebra / 745 & $5.956$ & $5.951$ & $0.999084$ & $5.933$ & $5.940$ & $1.001095$\\
\bottomrule
\end{tabularx}
\par\smallskip
\fontsize{8}{9.6}\selectfont\raggedright
\textit{Note.} Each time sums four sessions; ratios use unrounded totals. Ratios exceed one for five roots under loop4 and ten under packed. Packed/loop4 is below one for every root under either cut policy. The registered aggregate direct effects remain ratios of totals, not the mean of these ratios. Full root names, all session times and unrounded values are in the editable data.
\par\endgroup
\end{table}
\begin{table}[tbp]\caption{Other historical natural LOCAL/WHOLE panels with their original units, clocks and estimators.}\label{tab:natural}% Maintained native LaTeX body | tab:natural | inherited values; current numbering is automatic.
\begingroup
\fontsize{8.7}{10.5}\selectfont
\setlength{\tabcolsep}{3.5pt}\renewcommand{\arraystretch}{1.03}
\noindent
\begin{tabularx}{\linewidth}{@{}>{\raggedright\arraybackslash}p{0.235\linewidth}>{\raggedright\arraybackslash}X>{\raggedright\arraybackslash}X@{}}
\toprule
Fixed historical panel & RTX historical Q8 & A5000 8B Q8, target-only\\
\midrule
Unit and clock & Four roots, three sessions; native service & Two conversations, two rounds; controller service plus reset\\
\midrule
LOCAL / WHOLE & $1.007294$ & $1.042196$\\
Cost or estimator & Geometric mean of 12 paired ratios. & WHOLE 36.385\,s; LOCAL 37.921\,s.\\
\bottomrule
\end{tabularx}
\par\smallskip
\fontsize{8}{9.6}\selectfont\raggedright
\textit{Note.} The A5000 estimator is a ratio of sums over eight episodes, with four per arm. Native calls fall from 1,072 to 1,070; total time rises by 1.535\,s, including 1.420\,s in the shared greedy-sampling stage. No population interval is inferred for either descriptive panel. These profiles differ from M1 and the selected 32B LOCAL transactions.
\par\endgroup
\end{table}

The historical RTX panel has four roots and three sessions (12 paired records). The A5000 target-only panel has two conversations and two rounds (four episodes per arm); its controller clock includes reset and the shared greedy-sampling stage. These descriptive panels retain their original estimates, separately from the selected 32B transactions and later confirmations.

\FloatBarrier
\subsection{The stopped B570 draft-placement pilot}
\label{supp:pilot}
\begingroup\looseness=-1
The pilot moves 112 draft main matrices and its LM head to Vulkan0, retaining a CPU-mapped input embedding. Its two predeclared roots have SD\_C/TO\_C ratios 1.90760485 and 2.83651421, with geometric ratio 2.32614451. Mean draft prefill, synchronization and proposal cost is 14.807324\,s, exceeding the 1.642558\,s target decode-to-verification saving (Table~\ref{tab:pilot}). This cost account belongs to the two-root placement test; the formal 24-root CPU-draft study has its own 2.512087 service ratio.\par
\endgroup

\begin{table}[htbp]\caption{The separate B570 draft-placement pilot: two roots, one round and a predeclared complete-service stop.}\label{tab:pilot}% Maintained native LaTeX body | tab:pilot | inherited values; current numbering is automatic.
\begingroup
\fontsize{8.7}{10.5}\selectfont
\setlength{\tabcolsep}{3.5pt}\renewcommand{\arraystretch}{1.03}
\noindent
\begin{tabularx}{\linewidth}{@{}>{\raggedright\arraybackslash}Xrrr@{}}
\toprule
Complete-service contrast & Code root & Math root & Geometric ratio\\
\midrule
SD\_C/SD\_A & $0.797052$ & $0.837066$ & $0.816814$\\
SD\_C/TO\_A & $1.417249$ & $2.367886$ & $1.831907$\\
SD\_C/TO\_C & $1.907605$ & $2.836514$ & $2.326145$\\
\bottomrule
\end{tabularx}\par\smallskip\noindent
\begin{tabularx}{\linewidth}{@{}>{\raggedright\arraybackslash}Xr@{}}
\toprule
Pilot-only arithmetic stage account & Mean (s)\\
\midrule
Draft prefill / sync / proposal & $14.807324$\\
Target decode-to-verify saving & $1.642558$\\
\bottomrule
\end{tabularx}
\par\smallskip
\fontsize{8}{9.6}\selectfont\raggedright
\textit{Note.} Roots: CodeContests/test/124 and MATH/counting\_and\_probability/train/398. One measured round; service includes normal teardown; no population interval is assigned. The lower rows are arithmetic stage means, not geometric effects or a complete additive service decomposition. The formal 24-root CPU-draft study is separate.
\par\endgroup
\end{table}

% Keep the A.8 heading with its complete opening paragraph.
\begin{samepage}
\subsection{Additional historical results and their scope}
\label{supp:implementation}
The historical 32B scalar-reference application is 14.00\% slower than target-only across four roots, while its speculative-path change reduces time by 9.61\%. A separate shared-activation scalar S3 panel gives a 1.5071\% reduction [1.2436\%, 1.7554\%] across 12 roots and two observed dates. Its B/C paths and target-only clock differ from those in the modern C1/C4 request study.\par
\end{samepage}

A B570 fixed-append study on SmolLM2-1.7B~\cite{smol} uses cache capacities of 1024 and 4096 slots (C1024/4096), query widths q2, q4 and q32, and three paired repeats per cell. At its $1+(q-1)$ physical-view crossings, LOCAL reduces time by 20.34--44.68\%. For appends that stay within one physical extent, LOCAL takes 0.43--3.74\% more time. Earlier Qwen8B/SmolLM2 q32 cells with $16+16$ crossings include slower LOCAL results (Table~\ref{tab:historical-geometry}). Each cell retains its capacity, prefix, requested outputs and clock.

\begin{table}[htbp]\caption{Earlier Vulkan geometry observations with their original fixed-cell summaries.}\label{tab:historical-geometry}% Maintained native LaTeX body | tab:historical-geometry | inherited values; current numbering is automatic.
\begingroup
\fontsize{8.7}{10.5}\selectfont
\setlength{\tabcolsep}{2.8pt}\renewcommand{\arraystretch}{1.03}
\noindent
\begin{tabular*}{\linewidth}{@{\extracolsep{\fill}}lrrrrrr@{}}
\toprule
Width & \multicolumn{2}{c}{Qwen C1024 dual} & \multicolumn{2}{c}{Qwen C4096 target-only} & \multicolumn{2}{c}{Smol C4096 dual}\\
\cmidrule(lr){2-3}\cmidrule(lr){4-5}\cmidrule(lr){6-7}
$q$ & \multicolumn{2}{c}{9 blocks} & \multicolumn{2}{c}{12 blocks} & \multicolumn{2}{c}{6 blocks}\\
 & Crossing & Inside & Crossing & Inside & Crossing & Inside\\
\midrule
$2$ & $0.821784$ & $0.991242$ & $0.885123$ & $0.999833$ & $0.809705$ & $1.011280$\\
$4$ & $0.761142$ & $0.998782$ & $0.882388$ & $1.011929$ & NM & NM\\
$8$ & $0.838017$ & $0.996086$ & $0.928937$ & $0.990164$ & $0.791084$ & $1.016390$\\
$16$ & $1.047401$ & $0.999807$ & $1.010061$ & $0.997051$ & $0.923358$ & $0.982127$\\
$32$ & $1.055036$ & $1.014255$ & $1.014741$ & $0.999329$ & $1.088956$ & $1.001169$\\
\bottomrule
\end{tabular*}
\par\smallskip
\fontsize{8}{9.6}\selectfont\raggedright
\textit{Note.} All entries are LOCAL/WHOLE time ratios. NM: not measured. Source-bound historical Vulkan frontier summaries (immutable identities in the artifact). Ratios are inverted from the original W/L direction; below one favors LOCAL and above one favors WHOLE. Smol q4 was not measured. q32 crossings include $16+16$ partitions; these cells differ from the later $1+(q-1)$ append experiment. Their original fixed-cell timing windows and block counts are retained, without population intervals or pooling across capacity, model or geometry.
\par\endgroup
\end{table}

Earlier Qwen8B, Qwen14B and Llama resident studies report 17.21\%, 25.54\% and 11.52\% reductions against their fastest measured target-only controls. They use composite speculative policies and their own family-adjusted intervals; two Llama roots are slower. Llama's short-worker total reverses the resident advantage, 368.822 versus 355.730\,s, after loading, checks and primers enter the lifetime.\par

\begin{samepage}
Another historical RTX q8 equal-work study saves 3.94--11.91\% for the same eight known rows. Live speculation with seven draft proposals (k7) costs 13.58--14.59\% more than with three (k3) at equal emitted-token counts. The k7 runs process more verification rows and draft proposals.\par
\end{samepage}

\begin{table}[htbp]\caption{Historical resident effects and the separate worker lifetime. Fastest resident and worker controls are selected within their own measured configurations.}\label{tab:historical-applications}% Maintained native LaTeX body | tab:historical-applications | inherited values; current numbering is automatic.
\begingroup
\fontsize{8.5}{10.5}\selectfont
\setlength{\tabcolsep}{2.6pt}\renewcommand{\arraystretch}{1.03}
\noindent
\begin{tabularx}{\linewidth}{@{}>{\raggedright\arraybackslash}Xrrrrrrr@{}}
\toprule
Profile & Roots & \multicolumn{3}{c}{Resident L-P-D4 / fastest M} & \multicolumn{3}{c}{Worker lifetime (s)}\\
\cmidrule(lr){3-5}\cmidrule(lr){6-8}
 &  & Ratio & 99\% low & 99\% high & Fastest M & L-P-D4 & $\Delta$\\
\midrule
Qwen8B & $16$ & $0.827922$ & $0.753870$ & $0.888346$ & $200.739$ & $168.973$ & $-31.766$\\
Qwen14B & $16$ & $0.744608$ & $0.651094$ & $0.817914$ & $662.546$ & $594.559$ & $-67.987$\\
Llama8B & $12$ & $0.884779$ & $0.790486$ & $0.956653$ & $355.730$ & $368.822$ & $+13.092$\\
\bottomrule
\end{tabularx}
\par\smallskip
\fontsize{8}{9.6}\selectfont\raggedright
\textit{Note.} Worker $\Delta$ is L-P-D4 minus its fastest measured M; positive adds time. Resident ratio compares composite L-P-D4 with the fastest measured target-only M in each study; each five-contrast family also retains all other fixed M controls and W-L4-D4 in the editable source. All studies have four sessions per root. Worker totals include loading, checks and primers and use their own fastest measured M. Qwen8B has fixed relative D4 order; its compound effect can include order effects. Llama has two slower resident roots (ratios 1.017741 and 1.031393), and its worker total is slower.
\par\endgroup
\end{table}

The historical studies retain their original analyses. RefBound provides their frozen summaries and source identities; the released scalar-record coverage is defined below.

\par\penalty0\relax
\begin{samepage}
\subsection{Numerical qualification, source identities and public access}
\label{supp:access}
The public companion, \href{https://github.com/leecdiang/RefBound/tree/de2192a57149150f4fea6811db2c3e8781563e2c}{RefBound at commit \texttt{de2192a5}}, contains P4, LOCAL and RTX runtime source, scalar timing records and reconstruction scripts. Its \href{https://github.com/leecdiang/RefBound/blob/58697de93d86fc39c105bdfb8d8c0fd2949f192c/reproduce/SOURCE_MAP.json}{source map} links released derivatives to historical input hashes. The delivery for this revision provides the manuscript source and a separate \texttt{evidence/qualification\_manifest.json}, which records scientific configurations, executable and library SHA-256 identities, object counts, comparison scopes and results. Complete vocabulary/KV payloads and raw CUPTI buffers are retained privately on the original machines.\par
\end{samepage}

Tables~\ref{tab:deployment-details}--\ref{tab:qualification-details} describe the deployments, workloads and historical qualifications; Table~\ref{tab:access} summarizes the final CUDA checks.

\begin{table}[H]\caption{Final device qualification on fixed inputs, with complete numerical objects and each arm's own continuation.}\label{tab:access}% Maintained native LaTeX body | tab:access | 2026-11-21 device evidence.
\begingroup
\fontsize{8.7}{10.5}\selectfont
\setlength{\tabcolsep}{3.5pt}\renewcommand{\arraystretch}{1.03}
\noindent
\begin{tabularx}{\linewidth}{@{}>{\raggedright\arraybackslash}p{0.43\linewidth}>{\raggedright\arraybackslash}X@{}}
\toprule
Fixed device qualification & Complete-object and execution evidence\\
\midrule
\textbf{RTX 5070 Ti, C1/C4}\newline One 757-token input; one request per arm; P4 off & CUPTI verifies q1 versus q4/q2 target kernels. Each arm supplies 75 intermediate active F16 KV objects; paired bytes and historical hashes match. Terminal state matches the independent reference; 37 own-successor links pass per arm.\\
\midrule
\textbf{A5000, LOCAL crossing}\newline R/SIMPLE/LOCAL; 253-token history, frozen proposals; C4 on, P4 off & 27 full F32 vectors and 15 active F16 KV objects. Candidates match independent q1 R in 18 vector and 10 KV comparisons; 12 retained-prefix checks bind each arm's own continuation. All 42 objects also match same-arm historical captures.\\
\midrule
\textbf{A5000, RefBound P4}\newline Seven arms: q32 target-only and live SD on/off, two independent q1 controls, q31 guard & Same release ELF in all five candidate arms. Both q32 on arms record 1,780 P4 dispatches; q31 falls back. Captures contain 729 active KV and 1,260 complete F32 objects; 2,125 byte comparisons pass for independent references, paired modes and own continuation.\\
\bottomrule
\end{tabularx}
\par\smallskip
\fontsize{8}{9.6}\selectfont\raggedright
\textit{Scope.} These runs measure dispatch and numerical/state agreement. The statistical panels retain their original timing samples. Section~\ref{supp:access} identifies the binaries and separates public reconstruction inputs from original-machine numerical payloads.
\par\endgroup
\end{table}

The final LOCAL run fixes Qwen3-32B-Q4\_K\_M, CodeContests/train/161 at cycle 37, the 253-token history, C1024/F16, C4 on, P4 off, frozen proposals, and FlashAttention/CUDA graphs off. Reference $R$ issues four q1 decoder calls, SIMPLE issues q3+q1, and LOCAL issues one q4 call. Their own successors use four q1 calls, q1+q3 and q4 as attention extents change from $[256,256,256,512]$ to $[256,512,512,512]$. The rebuilt candidate and reference ELF identities are \texttt{67b9ce297e55} and \texttt{79de9f3740ef}; Table~\ref{tab:access} gives the full-object and continuation comparisons.

The RTX pair uses the same capture executable (\texttt{1e4c2a5c6955}) and unchanged CUDA backend DLL (\texttt{17b8a234f512}) on CodeContests/test/124. CUPTI associates 246,252 kernels per arm with request and phase scopes, with zero dropped or unassociated request records. Section~\ref{sec:implementation} reports the verification launches; Table~\ref{tab:access} records intermediate KV, reference and continuation agreement. The separate 864-request latency estimates remain in Section~\ref{supp:rtx-confirmation}.

% Place the qualification summary before the detailed P4 evidence.
\FloatBarrier
The A5000 P4 run uses one unmodified Linux/sm86 release ELF (\texttt{ac843f7b5e40}) built from the fixed RefBound source, with independent q1 ELF \texttt{aa86d1b1db1b}. Its five candidate arms share that binary: target-only and live SD with P4 off/on, plus a physical q31 guard control. P4 dispatch occurs during target prefill in the two q32 arms with P4 enabled. Arms with P4 off and other phases record zero. The 1,260 F32 objects comprise 1,253 rows and seven terminal vectors. The 2,125 byte comparisons comprise 1,396 across arms, 720 retained-prefix checks and nine observed/worker terminal checks. Both live SD arms execute 41 rounds, including 16 rejections, and match the independent q1 same-proposal replay. Read-only observation captures bytes already serialized by the unchanged ELF. The q31 arm checks guard fallback and its own state contract, with no same-input independent q31 comparison arm. This release ELF is qualified at q32; the historical q16/q24 tests in Table~\ref{tab:qualification-details} bind their earlier~binaries.

\par\penalty0\relax
\begin{samepage}
\paragraph{Statistical reconstruction.}
The public \href{https://github.com/leecdiang/RefBound/blob/de2192a57149150f4fea6811db2c3e8781563e2c/reproduce/README.md}{statistics entry point} reads 96 A5000 arm observations, 864 RTX formal observations, the 1,764-row RTX role register, 1,134 earlier contrast-level paired rows, and 256 measured M1 requests plus 32 primers. It reconstructs A5000 whole-block and RTX crossed root/block intervals, 27 point estimates from prior studies, 474 per-root point estimates, and the M1 points and interaction. Earlier-study and M1 intervals remain frozen metadata. Contrast rows can share request identities. The \href{https://github.com/leecdiang/RefBound/blob/de2192a57149150f4fea6811db2c3e8781563e2c/reproduce/SCOPE.md}{release scope} lists omitted identity fields and archives.\par
\end{samepage}

\paragraph{Manuscript and source reconstruction.}
The accompanying source package builds the present manuscript. The public \href{https://github.com/leecdiang/RefBound/blob/58697de93d86fc39c105bdfb8d8c0fd2949f192c/paper/build.py}{paper build} reconstructs the earlier released manuscript, figures and native tables from derivatives and frozen sidecars. The public \href{https://github.com/leecdiang/RefBound/blob/58697de93d86fc39c105bdfb8d8c0fd2949f192c/paper/DATA_MINIMIZATION.json}{figure-data register} and statistics companion select different fields from the historical records. The \href{https://github.com/leecdiang/RefBound/blob/58697de93d86fc39c105bdfb8d8c0fd2949f192c/runtime/PROVENANCE.json}{runtime register} identifies the released A5000 P4 core and application frontend. The \href{https://github.com/leecdiang/RefBound/tree/de2192a57149150f4fea6811db2c3e8781563e2c/runtime/local}{LOCAL diagnostic frontend} adds its build recipe, trace checks and saved-capture reader. Its new ELF compiled, linked and passed model-free checks but has no GPU qualification; the fixed-case numerical evidence above binds the original qualification binary. The released RTX C1/C4 capture worker was rebuilt and passed \href{https://github.com/leecdiang/RefBound/blob/de2192a57149150f4fea6811db2c3e8781563e2c/runtime/rtx/records/device-replay.json}{device verification}, without adding performance samples.

GPU replay requires fixed inputs and model weights; A5000 also needs the independent q1 executable, while RTX checks against saved independent reference objects. Complete numerical payloads remain on the original machines.

% Maintained native LaTeX register; data inherited unchanged from PR #290.
\begingroup
\fontsize{8.7}{10.5}\selectfont
\setlength{\tabcolsep}{3.5pt}\renewcommand{\arraystretch}{1.03}
\setlength{\LTleft}{0pt}\setlength{\LTright}{\fill}
\begin{longtable}{@{}P{.16\linewidth}P{.27\linewidth}P{\dimexpr.57\linewidth-4\tabcolsep\relax}@{}}
\caption{Deployment profiles and execution settings. Historical build and model identities specify the measured configurations; public access is defined in Section~\ref{supp:access}.}\label{tab:deployment-details}\\
\toprule
Study profile & Target / draft & Placement and execution settings\\
\midrule
\endfirsthead
\caption[]{Deployment profiles and execution settings (continued).}\\
\toprule
Study profile & Target / draft & Placement and execution settings\\
\midrule
\endhead
\midrule
\multicolumn{3}{r@{}}{\footnotesize Continued on the next page.}\\
\endfoot
\endlastfoot
\multicolumn{3}{@{}l}{\textbf{NVIDIA RTX A5000 / CUDA}}\\*
Modern & Qwen3-32B Q4\_K\_M / 0.6B Q8\_0 & CUDA, same GPU residence. Target microbatch/output reserve 32; draft microbatch 4. LOCAL reuses the prefill numerical core with P4 off.\\
\addlinespace[2.5pt]
M1 & Qwen3-8B / 0.6B Q8\_0 & Historical CUDA reference. Target microbatch 32/output reserve 4; draft microbatch 4. Layer placement NR.\\
\addlinespace[2.5pt]
Confirmation A & 32B / resident 0.6B & original LOCAL core, C1024; target 65/65 and draft 29/29 layers on GPU.\\
\addlinespace[2.5pt]
Confirmation B & 8B target-only & original natural-service core, C4096; target 37/37 GPU layers.\\
\midrule
\multicolumn{3}{@{}l}{\textbf{GeForce RTX 5070 Ti Laptop / CUDA}}\\*
Modern & Llama-3.1-8B-Instruct Q8\_0 / Llama-3.2-1B-Instruct Q8\_0 & CUDA; target 33/33 and draft 17/17 layers on GPU; both microbatch 32. Initial bridge and confirmation share binaries and numerical core.\\
\midrule
\multicolumn{3}{@{}l}{\textbf{Intel Arc B570 / Vulkan}}\\*
Formal & Llama 8B / 1B Q8\_0 & Vulkan F32-by-Q8 reference; target 33/33 GPU layers, draft 0/17. Target microbatch/reserve 32; CPU draft microbatch 1.\\
\addlinespace[2.5pt]
Pilot & Same models as formal & Separate placement test: 112 draft matrices and LM head on Vulkan0, embedding CPU-mapped; draft microbatch 32.\\*
\bottomrule
\noalign{\unskip\nobreak\vskip\belowbottomsep}
\multicolumn{3}{@{}p{\linewidth}@{}}{\fontsize{8}{9.6}\selectfont\raggedright \textit{Common settings.} Dense F16 KV and greedy decisions are common. CUDA profiles disable FlashAttention and CUDA graphs; RTX B/C1/C4 also disables fusion/PDL. Live SD uses three proposals and cap128. LOCAL uses frozen proposals and zero live draft forward work. Historical deployment records bind build, compiler, driver, device-clock, CPU-affinity, and model-byte identities; Section~\ref{supp:access} specifies their public access. NR denotes a field not recorded in the inspected historical bundle.}\\
\end{longtable}
\endgroup

% Keep the complete sampling register and its notes together.
\clearpage
% Maintained native LaTeX register; data inherited unchanged from PR #290.
\begingroup
\fontsize{8.7}{10.5}\selectfont
\setlength{\tabcolsep}{3.0pt}\renewcommand{\arraystretch}{1.03}
\setlength{\LTleft}{0pt}\setlength{\LTright}{\fill}
\begin{longtable}{@{}P{.197\linewidth}P{.11\linewidth}P{.35\linewidth}>{\raggedleft\arraybackslash}p{.11\linewidth}>{\raggedleft\arraybackslash}p{.10\linewidth}>{\raggedleft\arraybackslash}p{\dimexpr.133\linewidth-10\tabcolsep\relax}@{}}
\caption{Study sampling, workloads and clocks. Counts exclude warmup and qualification; footnotes define dependent successors and episode units.}\label{tab:study-details}\\
\toprule
Study & Sampling & Arms & Input & Output & Meas.\\
\midrule
\endfirsthead
\caption[]{Study sampling, workloads and clocks (continued).}\\
\toprule
Study & Sampling & Arms & Input & Output & Meas.\\
\midrule
\endhead
\midrule
\multicolumn{6}{r@{}}{\footnotesize Continued on the next page.}\\
\endfoot
\endlastfoot
\multicolumn{6}{@{}l}{\textbf{Selected transactions / verify through commit}}\\*
A5000 LOCAL & $1\times2\times2^{a}$ & SIMPLE, LOCAL & 137/253 & 4 rows & $8$\\
A5000 confirm A & $1\times2\times12^{b}$ & SIMPLE, LOCAL & 137/253 & 4 rows & $48$\\
\midrule
\multicolumn{6}{@{}l}{\textbf{CUDA request studies / reset through final synchronization}}\\*
A5000 C4 & $12\times2$ & C1, C4 & 39--187 & $128$ & $48$\\
RTX C4 initial & $24\times2$ & B, C1, C4 & 140--757 & NR & $144$\\
RTX confirm & $24\times12^{c}$ & B, C1, C4 & 140--757 & $42$--$128^{d}$ & $864$\\
A5000 P4 & $12\times4$ & TO/SD $\times$ C4/P4 & 39--187 & $128$ & $192$\\
RTX P4 & $24\times2$ & TO/SD $\times$ C4/P4 & 140--757 & 42--128 & $192$\\
\midrule
\multicolumn{6}{@{}l}{\textbf{Natural service / native-service clock}}\\*
A5000 M1 & $16\times4$ & W/L $\times$ loop4/packed & 55--790 & 59--128 & $256$\\
\midrule
\multicolumn{6}{@{}l}{\textbf{Natural service / two-turn response plus reset}}\\*
A5000 confirm B & $2\times12^{b}$ & SIMPLE, LOCAL & 27--551 & $2\times128$ & $48$\\
\midrule
\multicolumn{6}{@{}l}{\textbf{Vulkan service / compute plus normal teardown}}\\*
B570 formal & $24\times2$ & TO/SD $\times$ A/C & 140--757 & 44--128 & $192$\\
B570 pilot & $2\times1$ & TO/SD $\times$ A/C & 163/757 & $128$ & $8$\\
\bottomrule
\multicolumn{6}{@{}p{\linewidth}@{}}{\fontsize{8}{9.6}\selectfont\raggedright \textit{Units.} Input is prompt tokens (history tokens for transactions); output is generated tokens (requested rows for transactions). Sampling is roots $\times$ repeats unless annotated. Meas. counts formal arm observations; warmup and qualification are excluded. $^{a}$One root, two histories, two paired repeats; each first transaction has a dependent successor (history 139/255). $^{b}$A5000 confirmation repeats are time blocks; B's units are two-turn episodes and inputs are full per-turn prompts before prefix reuse. $^{c}$Six time blocks, two rounds per root/block. $^{d}$Visible output excludes EOG; decision sequences have 43--128 tokens. Modern application roots are balanced code/math; M1 has eight APPS and eight MATH roots. NR: unreported visible output range.}\\
\end{longtable}
\endgroup

% The complete S22 register fits on one page; avoid a one-row start.
\clearpage
% Historical qualification register; current device identities are in tab:access.
\begingroup
\fontsize{8.7}{10.5}\selectfont
\setlength{\tabcolsep}{3.5pt}\renewcommand{\arraystretch}{1.03}
\setlength{\LTleft}{0pt}\setlength{\LTright}{\fill}
\begin{longtable}{@{}P{.43\linewidth}P{\dimexpr.57\linewidth-2\tabcolsep\relax}@{}}
\caption{Historical numerical qualification: exercised cases, captured objects and coverage.}\label{tab:qualification-details}\\
\toprule
Bound path and exercised cases & Object and observation boundary\\
\midrule
\endfirsthead
\caption[]{Historical numerical qualification: exercised cases, captured objects and coverage (continued).}\\
\toprule
Bound path and exercised cases & Object and observation boundary\\
\midrule
\endhead
\midrule
\multicolumn{2}{r@{}}{\footnotesize Continued on the next page.}\\
\endfoot
\endlastfoot
\multicolumn{2}{@{}l}{\textbf{Original path qualifications}}\\*
\textbf{A5000 32B; C1/C4}\newline Six sentinel captures; q2--4 selector & Complete conditional vocabulary rows; active state and continuation.\\
\addlinespace[2.5pt]
\textbf{A5000 32B; R/C4/P4}\newline Independent q16/q24 panel; base32/base248 & 264 vocabulary and 36 state objects; crossings $8+8$ and $8+16$.\\
\addlinespace[2.5pt]
\textbf{A5000 32B; P4}\newline 48 captured warm requests, 12 roots $\times$ four arms & Full requested vectors and committed checkpoints; application P4 calls q32.\\
\addlinespace[2.5pt]
\textbf{A5000 32B; LOCAL}\newline Two histories, first transaction and own successor & Owned complete rows, tentative/committed KV, pending token and continuation.\\
\addlinespace[2.5pt]
\textbf{RTX 8B/1B; C4/P4}\newline q4/q32 and registered application qualification & Complete vocabulary, active KV and continuation.\\
\midrule
\multicolumn{2}{@{}l}{\textbf{Confirmation studies}}\\*
\textbf{RTX B/C1/C4 confirmation}\newline Two sentinel roots, 24 included qualification requests & New candidate captures against stored reference vectors/KV; timed-request complete-state/frontier and actual-work bindings.\\
\addlinespace[2.5pt]
\textbf{A5000 confirmation}\newline Same 32B histories; original 8B two-turn consumer; 12 pre-timing qualification requests & Full owned/frontier vectors, canonical active state, commit and own continuation; capture counts in Section~\ref{supp:confirmation}.\\
\bottomrule
\multicolumn{2}{@{}p{\linewidth}@{}}{\fontsize{8}{9.6}\selectfont\raggedright \textit{Coverage.} These historical qualifications bind their original binaries and use dense nonwrapping F16 KV, greedy decisions and non-FlashAttention. The q16/q24 row applies to the historical A5000 panel; Table~\ref{tab:access} records the later release ELF separately. Separate q12/q20/q28 and complete C4 logical-tail q1/EOG captures are absent; RTX has no complete q16/q24 qualification.}\\
\end{longtable}
\endgroup

\clearpage
\subsection{Related numerical and state interfaces}
\label{supp:related-contracts}

Table~\ref{tab:neighbors} records the fixed references, execution choices and state interfaces supporting the comparisons in Section~\ref{sec:related}. LOCAL's row describes the construction in Figure~\ref{fig:construction} under its stated operator and state conditions.

\begin{table}[!htbp]\caption{Numerical references and execution/state contracts in related systems. LOCAL describes the construction in Figure~\ref{fig:construction}.}\label{tab:neighbors}% Table 7 | tab:neighbors | editable body generated by scripts/build_tables.py
\begingroup
\fontsize{8.7}{10.5}\selectfont
\setlength{\tabcolsep}{4.0pt}\renewcommand{\arraystretch}{1.03}
\noindent
\begin{tabularx}{\linewidth}{@{}>{\raggedright\arraybackslash}p{0.31\linewidth}>{\raggedright\arraybackslash}p{0.32\linewidth}>{\raggedright\arraybackslash}X@{}}
\toprule
System and fixed reference & Execution choices & State and consumer\\
\midrule
\textbf{Taming~\cite{taming}}\newline Compiled reference: first configuration or caller-supplied kernel & Layout and launch within a verified structural class & Studied contract is kernel-level.\\
\addlinespace[2.5pt]
\textbf{Detent~\cite{detent}}\newline Fixed model, deployment and commitment tuple & Dependency cuts identify pins; legal pin values and deployment knobs are tuned before serving & Serving, rollout and training use the committed tuple.\\
\addlinespace[2.5pt]
\textbf{Vosti~\cite{vosti}}\newline Canonical prefix values under a fixed deployment & Scheduling, query lengths and physical KV mappings & Verified engine and kernel contracts preserve KV, write isolation and continuation.\\
\addlinespace[2.5pt]
\textbf{BlockLock~\cite{blocklock}}\newline Canonical Rows/Pack geometry, length class and batch size & Companion requests vary around protected placement & Batched Rows or target-only Pack decode; exact KV and position handoff.\\
\addlinespace[2.5pt]
\textbf{LLM-42~\cite{llm42}}\newline Fixed-shape verification path & Dynamic fast path; verification groups can share a batch & Verified prefix commits; verifier KV replaces fast-path KV.\\
\midrule
\textbf{This work}\newline Inherited projection recipe, attention views and consumer & Retain q4 around the local $3+1$ attention refinement & Own ordered rows; commit active KV and pending token; guard the arm's own successor.\\
\bottomrule
\end{tabularx}
\par\endgroup
\end{table}

\paragraph{Implementation and reading scope.}
Vosti's attention kernel preserves state on extra fully masked tiles~\cite{vosti}. LLM-42 combines a dynamic fast path with fixed-shape verification. Verifier groups can share a batch, and verifier KV replaces fast-path KV even on token agreement~\cite{llm42} (\href{https://www.microsoft.com/en-us/research/wp-content/uploads/2026/03/2601.17768v2.pdf}{author version, Sections 4.2--4.3}). QSpec replaces accepted draft KV with higher-precision verified KV~\cite{qspec}. QuantSpec combines hierarchical KV quantization with recent full-precision rejection buffers~\cite{quantspec}. MarginGate targets current-column KV repair after margin-triggered verification; its prototype uses rollback re-decoding~\cite{margingate}.

ESPEON's design transfers prefix KV for suffix checks and commits hashes of recorded states before verifier challenges; its evaluated prototype implements full verification~\cite{espeon}. These verification scopes and commitments differ from LOCAL's operator partitions and consumer commit.

\paragraph{Kernel and deployment interfaces.}
Tensor Core models describe architecture-specific rounding and normalization~\cite{tensorcoremodels}. FlashAttention and FlashInfer optimize attention layouts and schedules~\cite{flashattention,flashinfer}. FlashInfer fixes aggregation order for given sequence-length metadata. FlashInfer's later \href{https://github.com/flashinfer-ai/flashinfer/blob/3db2f474e4053a6400d3128ba0683cbd4a05f5a4/flashinfer/batch_invariant/README.md}{pinned documentation} defines one BF16 attention class per head dimension across prefill and decode with paged or ragged KV~\cite{flashinferbi}. The inspected vLLM paths combine \href{https://github.com/vllm-project/vllm/blob/01fb22eed6b5ae67f7b97bc19dc637bd413851c7/vllm/model_executor/determinism/batch_invariant_configs.py}{fixed reduction tiles with query-count-dependent output tiles and launches}. The AWQ branch \href{https://github.com/vllm-project/vllm/blob/01fb22eed6b5ae67f7b97bc19dc637bd413851c7/vllm/model_executor/layers/quantization/auto_awq.py\#L895-L911}{dequantizes before matrix multiplication}~\cite{vllm_batchinv}. CoRun isolates prefill and fixes decode shapes~\cite{corun}. Tritify separates its exact-integer reproducibility from decision fidelity to a floating-point reference~\cite{tritify}.

\FloatBarrier
\label{page:supplement-end}
\end{document}